\documentclass[12pt,a4paper]{report}

\usepackage{epcc}
\usepackage{graphics}
\usepackage{graphicx}
\usepackage{amsmath}
\usepackage{amssymb}
\usepackage{upgreek}
\usepackage{cite}
\usepackage{refstyle}
\usepackage{caption}
\usepackage{subcaption}
\usepackage{float}
\usepackage{fullpage}
\usepackage{tikz}
\usepackage{tkz-berge}
\usepackage{listings}
\usepackage{commath}
\usepackage[toc,page]{appendix}

\definecolor{keywords}{RGB}{255,0,90}
\definecolor{comments}{RGB}{0,0,113}
\definecolor{red}{RGB}{160,0,0}
\definecolor{green}{RGB}{0,150,0}

\graphicspath{{Images/}}

\begin{document}

\title{Blocking Vs. Non-Blocking Halo Exchange for Lattice Boltzmann}
\author{Anthony Bourached}

\date{}

\makeEPCCtitle

\begin{center}

    \includegraphics[width=0.5\textwidth]{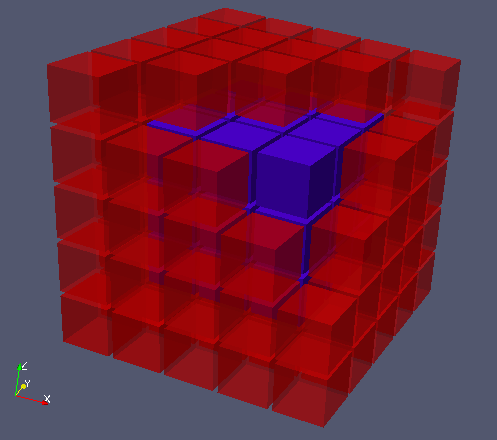}
	        
    \thispagestyle{empty}

    \vspace{3cm}

	\large 
	MSc in High Performance Computing\\
	The University of Edinburgh\\
	Year of Presentation: 2016\\
	19 August 2016
\end{center}

\newpage

\pagenumbering{roman}

\tableofcontents

\listoffigures

\pagenumbering{arabic}

\newpage

\begin{abstract}
	This report describes the design, implementation and analysis of a non-blocking halo exchange routine as an alternative to the blocking halo exchange routine in the lattice Boltzmann code Ludwig. The alternative non-blocking routine is implemented in such a way to allow work-communication overlap. Detailed benchmarks in this report show that the non-blocking version is a good alternative with some advantages and no significant disadvantages to the blocking version even without any work-communication overlap. Work-Communication overlap can be used to improve the performance of the non-blocking routine. Development and benchmarking were conducted on the UK national supercomputer, ARCHER.
\end{abstract}

\chapter{Introduction and Theory}
	 
\paragraph{}
Computer simulation is a massive and rapidly growing application in science. Processing power increases exponentially with time, however we can also greatly extend what we can simulate by carefully considering the efficiency, or performance, of our program. Therefore correct code is far from our only consideration when writing a program. There are many ways to perform any function and some are a lot better than others. This can be for reasons of performance, readability or robustness. One of the major focuses of the subject of High Performance Computing (HPC) is that of parallel programming.

\paragraph{}
HPC generally refers to the practice of aggregating computing power in a way that delivers much higher performance than one could get out of a typical desktop computer. A powerful modern desktop computer typically has eight cores. The purpose of multiple cores is to maximise efficiency of the computer's compute power. It allows the computer to more effectively perform different tasks at the same time. Maximum program efficiency would be achieved if the program's tasks were divided in such a way that they can be executed by independent cores with, preferably, minimum communication or co-dependency. This implies that the goals of HPC require that both software and the hardware involved in completing the tasks be considered carefully. 

\paragraph{}
One of the major applications of HPC in modern science is the ability to simulate an environment which implements the laws of physics virtually within it. In this way, experiments can be conducted in situations that would be impossible, or certainly inplausible, to set up in reality. Furthermore, the aggregated computing power of parallel programming enables the simulation of very complex environments such as modelling large DNA structures, global weather forecasting, modelling motion of astronomical bodies in space or the velocities of billions of atomic sized particles in a gas \cite{parallel_programming}. Conclusions can be reached from such simulations that could not be derived analytically.  

\paragraph{}
Lattice gas techniques have gained huge popularity over the past three decades \cite{IBM}, particularly owing to their ideal candidacy for massive parallel computing and the rapidly growing use of parallelism in high-end computing. Furthermore, it has the greatest flexibility in terms of handling complex geometries than any other kind of technique \cite{Yuanxun}.

\paragraph{}
In this report we consider a code developed at Edinburgh Parallel Compute Centre (EPCC), at the University of Edinburgh, called Ludwig \cite{ludwig}. This code is written in C and uses the Lattice Boltzmann Equation (LBE) \cite{Succi}, or, more directly, the Lattice Boltzmann Method (LBM) \cite{Yuanxun} to solve fluid-dynamic problems. This fully operational code can scale to very large problems and work efficiently in parallel on a large number of cores. In section \ref{sec:boltz_theory} we shall discuss the theory and background of the LBM and how it is employed in the Ludwig code before discussing the focus of our research interest and the underlying motivation in the further development of this code in sections \ref{sec:domain_decom}, \ref{sec:blockingVnonblock} and \ref{sec:mpi_message_cost}.

\paragraph{}
In Chapter \ref{chapter:development} we shall take a look at the development process and the structure of the code, specifically, the  Non-Blocking (see section \ref{sec:blockingVnonblock}) version. This description will include the code structure and design, the development process, considerations taken to maximise efficiency as well as tests conducted to ensure correct functionality of the developed code.

\paragraph{}
Chapter \ref{chapter:methodology} shall discuss the condition under which the benchmarks were conducted and the data analysed. From both a methodological and hardware perspective.

\paragraph{}
Chapters \ref{chapter:1_node} and \ref{chapter:strong_scaling} shall largely present results and analysis of the proficiencies of the two approaches to halo exchange (see section \ref{sec:blockingVnonblock}) with the aim of making conclusions that generalise to all large scale lattice models rather than being restricted to just Ludwig or other LBM codes. 

\paragraph{}
Chapter \ref{chapter:overlap} shall illustrate the potential advantage of the non-blocking version in terms of work-communication overlap (see section \ref{sec:non_blocking_halo_exchange}). This overlap is specific to certain problems in the Ludwig code. However, it gives the reader a strong indication of the potential benefits of the non-blocking version.

\paragraph{}
Finally, in Chapter \ref{chapter:conclusions} the overall achievements and observations of this project will be summarised. The most important discovery will be highlighted and its implications underlined. Furthermore, this project paves the way for further work which shall also be discussed in this chapter. This will conclude the report, however, an appendices is also provided.

	\section{The Lattice Boltzmann Method (LBM)} \label{sec:boltz_theory}

	\paragraph{}
	Gases and fluids can be considered as consisting of a large number of small particles moving in random directions. In this way the transfer of momentum and energy is achieved in the interaction of these particles \cite{Yuanxun}.
	
	\paragraph{}
	The Boltzmann equation \cite{Succi} is a statistical description of a particle in a fluid. The concept of the probability of a particle to be in a location in space, $\vec{r}$, at a given time, $\vec{t}$, is represented by a density distribution (or particle distribution function) $f(r,t)$. The Boltzmann transport equation is given by
	
	\begin{align}
		\frac{\partial f}{\partial t} + \vec{u} \cdot \nabla f = \Omega,
		\label{eq:boltz_eq}
	\end{align} 
	
	where $\vec{u}$ is the particle velocity and $\Omega$ is the collision operator which is a function of the particle distribution function and represents the rate of change of $f$ resulting from the collision. The LBM reformulates the idea of randomly distributed particles to particles confined to a lattice node. Which, instead of themselves propagating, exchange values of momentum and energy with their nearest neighbours. Thus, we can consider the LBM as a discretised representation of the LBE.
	
\begin{figure}[H]
		\centering
		\includegraphics[width=0.5\textwidth]{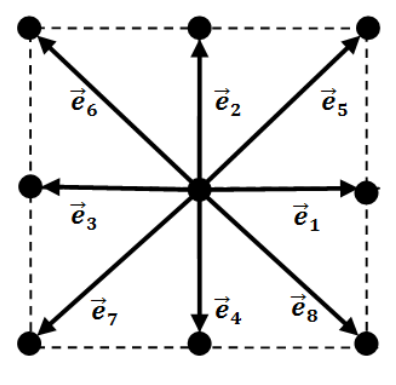}
		\caption{$D2Q9$. Microscopic velocities from each lattice point propagate to each of the 8 nearest neighbours on a square lattice. $\vec{e_0}$ remains on the original lattice site.}
		\label{fig:D2Q9}
	\end{figure}

	\paragraph{}
	For simplicity we shall first consider a 2-dimensional discretised lattice. For each observable\footnote{A variable of interest which quantifiably describes some property of the system.} considered in the simulation (such as energy, velocity, momentum etcetera) at each lattice site there are 9 \textit{microscopic observables} which each represent a component of that observable which shall propagate to each of the nearest neighbours, and itself. For the 2D case there are 9 \textit{microscopic velocities} ($\vec{e_{i}}$,   i = 0,..., 8) for each lattice site. This is illustrated in figure \ref{fig:D2Q9}.

	\paragraph{}
	We differentiate between different lattice Boltzmann models using $DnQm$ notation, where $n$ is the number of dimensions and $m$ is the number of discrete velocities allowed, including the 'on site'\footnote{Meaning the original lattice site.} velocity. Now $f_i(\vec{r}, \vec{e_i}, t)$ describes the probability of propagating in a particular direction.
		
	\paragraph{}
	As collisions only occur on lattice sites, the discretisation of space and time means that propagations and collisions can be computed separately, at two distinct stages of each iteration\footnote{An iteration consists of one, and only one update of the entire lattice in the simulation. This corresponds to a unit of discretised time.}. So the LBM can be described by
	
	\begin{align}
		f_i(\vec{r} + \vec{e_i}, t + 1) - f_i(\vec{x}, t) = - \frac{\abs{f_i(\vec{r}, t) - f_i^{eq}(\vec{r}, t)}}{\tau},
		\label{eq:boltz_discretised}
	\end{align}
	
	where $f_i^{eq}(\vec{r}, t)$ is the distribution required for equilibrium and $\tau$ is the relaxation time toward \textit{local equilibrium.} The left hand side of this equation represents the  propagation of particles through the lattice each time step by their microscopic velocities $\vec{e_i}$. The right hand side describes the systems approach toward equilibrium due to collisions.
	
	\paragraph{Macroscopic properties} of the fluid can be determined simply from the microscopic values. The \textit{macroscopic fluid density} is given by
	
	\begin{align}
		\rho (\vec{r}, t) = \sum_{i=0}^{m - 1} f_i(\vec{r}, t)
		\label{eq:macro_density}
	\end{align}
	
	where $m$ is number of discrete velocities allowed for the model and $\rho$ is the macroscopic fluid density. Similarly, the \textit{macroscopic velocity} can be given by 
	
	\begin{align}
		\vec{u} (\vec{r}, t) = \dfrac{1}{\rho(\vec{r}, t)}\sum_{i=0}^{m - 1} f_i(\vec{r}, t) e_i
		\label{eq:macro_velocity}
	\end{align}
	
	where $\vec{u}$ is the macroscopic velocity.
	
		\subsection{The Problem of Problem Size}
		
\begin{figure}[H]
			\centering
			\includegraphics[width=0.5\textwidth]{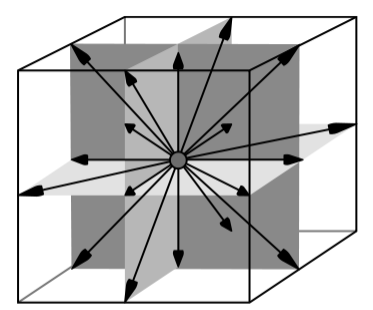}
			\caption{$D3Q19$. Microscopic velocities from each lattice point propagate to each of the 18 nearest neighbours on a cubic lattice. The 19th discrete velocity is 'on site'.}
			\label{fig:D3Q19}
		\end{figure}

		\paragraph{}
		The model of Ludwig with which we are concerned is $D3Q19$ as depicted in figure \ref{fig:D3Q19}. However our work is generalised such that the timings and the performance are valid for any model with up to 27 discrete velocities. In order to run a relatively standard size lattice with dimensions $512 \times 512 \times 512$ (134,217,728 sites) the amount of memory required is
		
		\begin{align}
			8 \times 19 \times 512^3 \approx 20 G Bytes,
			\label{eq:ram}
		\end{align}
		
		since for each lattice site we have 19 micro velocities- each a floating point number taking up 8 bytes of memory. 20 GigaBytes is more Random Access Memory (RAM) than can practically be provided by a single computer core or any shared memory environment. Hence the parallelism enabled by using such a lattice gas model is essential to form a proficient simulation. 
	
	\section{Domain Decomposition in Ludwig} \label{sec:domain_decom}
	
\begin{figure}[H]
		\centering
		\includegraphics[width=0.9\textwidth]{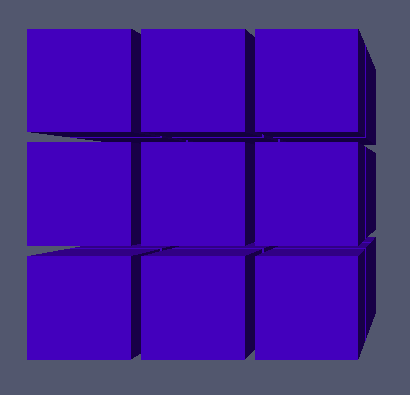}
		\caption{A domain decomposition of a cubic lattice. One domain has been split into $3 \times 3 \times 3 = 27$ subdomains. This corresponds to 27 MPI tasks, one iterating each subdomain.}
		\label{fig:domaindecom}
	\end{figure}

	\paragraph{}
	Ludwig is a practical and efficient domain decomposition (DD) algorithm which employs the LBM to iterate the solution on each subdomain. Domain decomposition is a typical parallel design pattern\footnote{Parallel Design Patterns are a set of ways of decomposing a problem to be solved in parallel- using multiple processors to solve a single problem.} for solving lattice models such as this which are a function of both space and time. This is  accomplished by splitting the lattice into smaller lattices on subdomains and iterating to coordinate the solution between adjacent subdomains. Ludwig uses the Message-Passing Interface (MPI) for parallel communications, where each subdomain is 'governed' by one MPI task which typically corresponds to one processor. An example of how this decomposition works for a cubic lattice is shown in figure \ref{fig:domaindecom}. Each subdomain seen in this figure is a lattice which uses LBM to iterate the values at its lattice sites.

	\paragraph{}
	In order to coordinate the solution between adjacent subdomains, and since this is a lattice problem (with nearest neighbour interaction), a certain amount of communication is required between adjacent subdomains after each iteration. Considering the central subdomain in figure \ref{fig:domaindecom}, like the centre of a rubic cube, it can be seen that it has 26 neighbours with which it must exchange its boundary lattice point values after each iteration. This is done by extending the dimension of the subdomain by one lattice point in each direction where this extra lattice point value is taken, or recevived, from the subdomain's neighbour in that direction.
	
	\begin{figure}[h]
		\centering
		\includegraphics[width=0.75\textwidth]{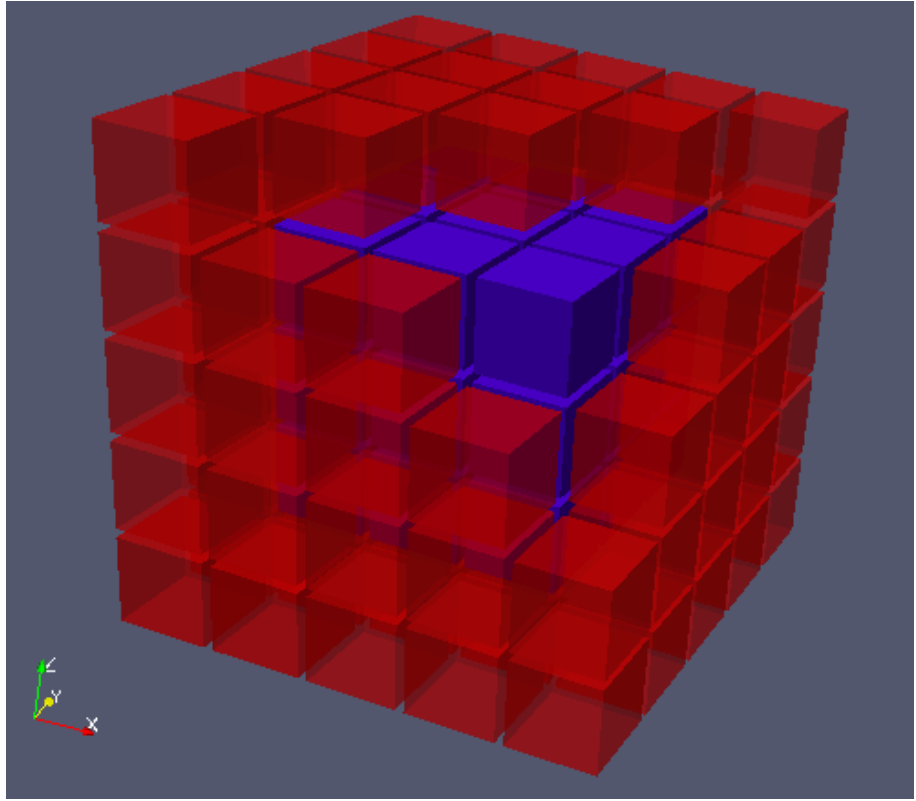}
		\caption{A single subdomain and its halo values. The blue lattice sites are part of the subdomain while the red, transparent lattice sites are the halo sites received from adjacent subdomains.  Note that a few halo sites belonging to the closest corner have been excluded for the purpose of illustration.}
		\label{fig:halo1}
	\end{figure}
	
	\paragraph{}
	 Figure \ref{fig:halo1} shows an example of a subdomain and its halo lattice points. In analogy to figure \ref{fig:domaindecom} the blue cubes represent lattice points (sites) and the set of blue cubes as a whole represent a single subdomain while the halo lattice sites received from neighbouring subdomains are transparent red. Each of these lattice sites, halo or non-halo, typically contains 19 floating point values (for the $D3Q19$ model) as discussed in section \ref{sec:boltz_theory}. Note that this subdomain is conceived for illustration purposes only; using a subdomain with so few lattice sites would be counter-productive due to the overhead of communication required to fill the halo sites. This shall be discussed in more detail later.

	\section{Blocking Vs. Non-Blocking Halo Exchange} \label{sec:blockingVnonblock}

		\subsection{Blocking Halo Exchange Implementation in Ludwig}
		\label{sec:blocking_halo_exchange}
	
		\paragraph{}
		The original implementation of the Ludwig halo exchange seeks to minimise the number of messages\footnote{A send and receive shall be referred to as one message.} sent, and received by each MPI task at each iteration necessary to complete each subdomains' halo buffers (as seen in figure \ref{fig:halo1}).
		
\begin{figure}[H]
			\centering
			\begin{subfigure}[h]{0.49\textwidth}
				\centering
				\includegraphics[width=\textwidth]{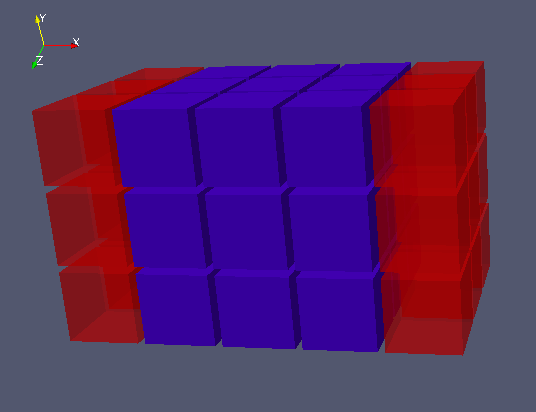}
				\caption{Red halos have been received in the x-direction (the yz plane).}
				\label{fig:sendX}
			\end{subfigure}
			\hfill
			\begin{subfigure}[h]{0.45\textwidth}
				\centering
				\includegraphics[width=\textwidth]{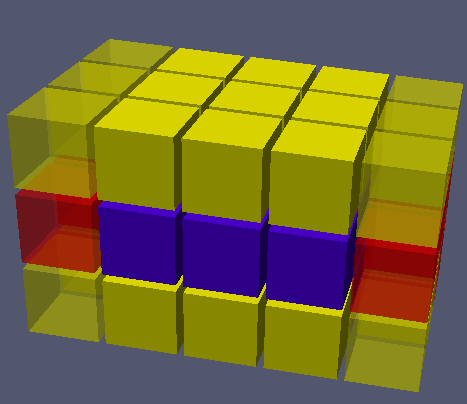}
				\caption{Highlighted yellow lattice sites are to be sent in y-direction (xz plane).}
				\label{fig:sendY}
			\end{subfigure} 
			\caption{Subdomain after communication in x-direction (yz plane) only. Sites highlighted yellow are to be sent next. Subdomain lattice size in this example is $3 \times 3 \times 3 = 27 $.}
			\label{fig:blocking_X}
		\end{figure}

		\paragraph{}
		Each subdomain needs to receive data from 26 neighbouring subdomains in order to have a complete halo buffer. It is possible however, to achieve this with only 6 messages (6 sends and 6 receives per subdomain). The first 2 messages are 2 sends (and correspondingly 2 receives) in the x-direction, i.e. the yz plane. Figure \ref{fig:sendX} shows the data present on each process after these 2 messages are received (one back and one forward in the x-direction). 
		
		\paragraph{}
		Figure \ref{fig:sendY} shows each process in the same state as figure \ref{fig:sendX}, however it highlights in yellow the data that is to be sent in the next 2 messages in the y-direction (the xz plane). Included in the halo values that are to be sent next are some that were received during the previous 2 messages, thus all communication in the x-direction needs to be finished before any data can be sent in the y-direction.
		
\begin{figure}[H]
			\centering
			\begin{subfigure}[h]{0.49\textwidth}
				\centering
				\includegraphics[width=\textwidth]{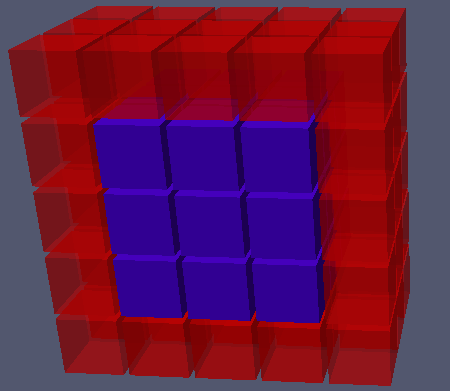}
				\caption{Red halos have been received in first 4 messages.}
				\label{fig:sentY}
			\end{subfigure}
			\hfill
			\begin{subfigure}[h]{0.45\textwidth}
				\centering
				\includegraphics[width=\textwidth]{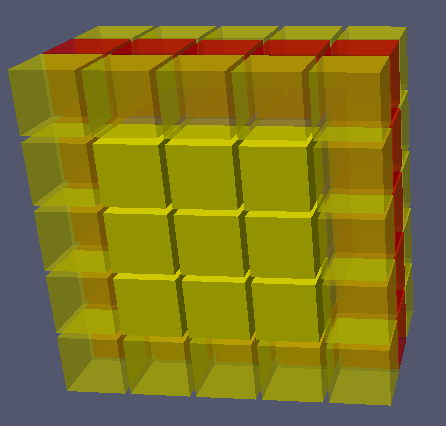}
				\caption{Highlighted yellow lattice sites are to be sent in z-direction (xy plane).}
				\label{fig:sendZ}
			\end{subfigure} 
			\caption{After messages have been sent and received in y-direction (xz plane). Sites highlighted yellow are to be sent next. Subdomain lattice size in this example is $3 \times 3 \times 3 = 27 $.}
			\label{fig:blocking_Y}
		\end{figure}

		\paragraph{}
		Once the highlighted data in figure \ref{fig:sendY} has been sent, and received, by each process; each subdomain appears as figure \ref{fig:sentY}. The top-edge halo sites (the ones diagonally adjacent to the lattice sites at the top edge of the subdomain in figure \ref{fig:sentY}) have each come from a diagonally adjacent subdomains since they have been translated once in the x-direction and once in the y-direction.

		\paragraph{}
		Finally, it is now clear to see how we obtain a complete set of halo data- as can be seen visually in figure \ref{fig:complete_halo}- by the final 2 sends and receives highlighted in figure \ref{fig:sendZ}.
	
		\paragraph{}
		This implementation of halo exchange minimises the number of messages sent however there is a significant trade off when having such a small number of messages: all subdomains must complete both of their sends and receives in the x-direction before any of them can begin their sends and receives in the y-direction. Effectively the simulation must wait for the last processor to complete its sends and receives in the x-direction before any processor may proceed, indubitably resulting in many MPI tasks being idle for a time. This is referred to as 'blocking'. Each iteration of the halo exchange process needs to block 3 times for a 3 dimensional lattice (once for each dimension as illustrated in figures \ref{fig:blocking_X} and \ref{fig:blocking_Y}). 
		
		\subsection{Non-Blocking Alternative} \label{sec:non_blocking_halo_exchange}

		\paragraph{}
		As we discussed in section \ref{sec:non_blocking_halo_exchange} each subdomain needs to communicate with 26 other subdomains each iteration. The only way to avoid the issue of blocking is for each subdomain to explicitly communicate with all of its neighbours. Thus requiring 26 sends and receives, rather than 6.
		
\begin{figure}[H]
			\centering
			\begin{subfigure}[h]{0.49\textwidth}
				\centering
				\includegraphics[width=\textwidth]{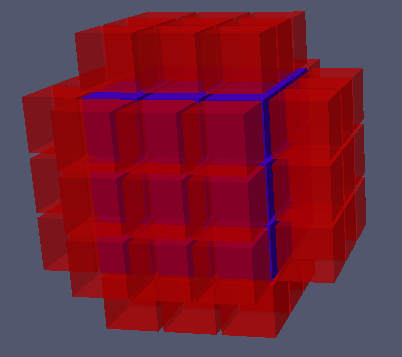}
				\caption{6 planes.}
				\label{fig:planes}
			\end{subfigure}
			\hfill
			\begin{subfigure}[h]{0.45\textwidth}
				\centering
				\includegraphics[width=\textwidth]{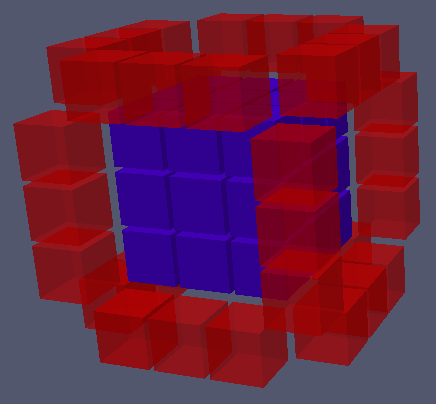}
				\caption{12 edges.}
				\label{fig:edges}
			\end{subfigure} 
			\begin{subfigure}[h]{0.45\textwidth}
				\centering
				\includegraphics[width=\textwidth]{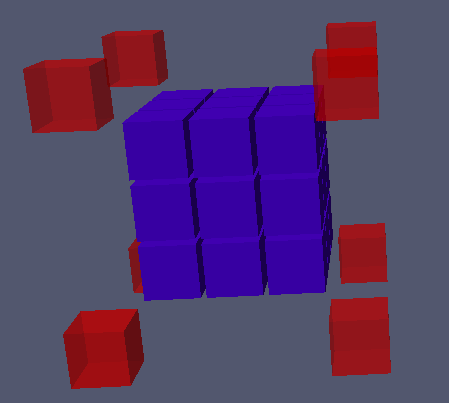}
				\caption{8 corners.}
				\label{fig:corners}
			\end{subfigure}
			\caption{Illustrating 26 messages involved in non-blocking halo receives, highlighted by red, transparent blocks. Subdomain lattice (blue) size in this example is clearly $3 \times 3 \times 3 = 27 $.}
			\label{fig:non_blocking}
		\end{figure}

		\paragraph{}
		Figure \ref{fig:non_blocking} illustrates the received halo data (in red): 
		
		\begin{enumerate}
			\item 6 receives contain the planes at the six faces of each subdomain, received from orthogonally adjacent subdomains (figure \ref{fig:planes}).
			\item 12 messages are received from subdomains diagonally adjacent to the edges (figure \ref{fig:edges}).
			\item Finally, eight messages are received from diagonally adjacent subdomains to the corners of the subdomain's lattice (figure \ref{fig:corners}).
		\end{enumerate}
		
		\paragraph{}
		This version needs to block once- when all communication has been completed- but it shall be referred to as the non-blocking version as it executes halo exchanges without the need to block. Furthermore, it is possible to include more work- as long as it is not dependent on the halo sites- before the blocking barrier but after messages have been sent. Concisely: this blocking version gives scope for further overlapping of work and communication. The focus of this report is the performance difference between the blocking and non-blocking version without any further overlap with the aim of seeing if the non-blocking version hinders performance it does not capitalise on the overlap advantage. Work-communication overlap, however, shall also be considered in chapter \ref{chapter:overlap}.
		
		\paragraph{}
		The total amount of bytes exchanged using each method is identical but there is an overhead cost with sending a greater number of messages as shall be discussed next.

		\subsection{Motivation for Developing Non-Blocking Version} \label{sec:motivation}
		
		\paragraph{}
		Davidson \cite{Davidson} improved the efficiency of the parallel communication in the Ludwig code by using MPI derived datatypes to reduce the amount of data sent between processors during each iteration. This was accomplished by only sending the microscopic velocities that would actually propagate into neighbouring subdomains. For example, in the $D3Q19$ (refer to figure \ref{fig:D3Q19}) model only five of the microscopic velocities need to be exchanged for each pair of orthogonally adjacent lattice sites on separate subdomains. This is a significant reduction in message size compared to sending nineteen floating point numbers for each boundary lattice site. 
		
		\paragraph{}
		Davidson found that this method of 'reduced' halo exchange communication improved the performance of the Ludwig code in all circumstances. This method used the same blocking communication as described in section \ref{sec:blocking}, as do all halo exchange routines in Ludwig. Thus it remains an outstanding challenge to determine the performance of a non-blocking halo exchange routine. Therefore, the aim of this project is to develop and benchmark a new version of the halo exchange routine that implements messages in the non-blocking way described in section \ref{sec:non-blocking}.
		
		\paragraph{}
		With a fully implemented non-blocking version of the halo exchange routine the trade-off between the advantage of sending few messages and the advantage of desynchronising (non-blocking) communication can be examined. Furthermore, the potential advantage of overlapping work and communication may be investigated. This halo exchange routine is not specific to Ludwig. Thus, the results of this project are applicable to any  lattice model simulations that have nearest neighbour interaction. This may be particularly significant for simulations that have a large portion of work that is not dependent on halo data since this can be overlapped with communication in the non-blocking version.

	\section{Cost of sending MPI messages} \label{sec:mpi_message_cost}
	
	\paragraph{}
	To determine the most efficient way to manage communications in largely parallel programs we must consider what variables effect the cost of communication. The time taken to send data from one processor to another using MPI messages is given by
	
	\begin{align}
		t = (l + \frac{m}{B}) \label{eq:mpi_cost},
	\end{align}
	
	where $l$ is Latency ($s$), the overhead involved in starting the message (regardless of message content, size or destination). $B$ is the bandwidth (MBytes/s), the amount of data that can be sent through the network from one point to another per second. Finally, $m$ is the size (MBytes) of the message. 
	
	\paragraph{}
	Given that the amount of data to be transferred in MPI communications for a given simulation is constant (i.e. $\sum_{1}^{N} m = M$ where N is total number of MPI messages and M is the total amount of data to be sent-which is constant for a given lattice size) performance of communication in parallel simulations is generally optimised by sending as few messages as possible. This can be seen mathematically by considering the total time for all MPI communications in a simulation:

	\begin{align}
		T =& \sum_{1}^{N} l + \sum_{1}^{N} \bigg(\frac{m}{B}\bigg) \label{eq:mpi_total_cost}, \\
		T =& \sum_{1}^{N} l +  M \label{eq:mpi_cost_minimised},
	\end{align} 
	
	where $M = \Big(\frac{\sum_{1}^{N}m}{\sum_{1}^{N}B}\Big)$ is constant for a given simulation size. Equation \ref{eq:mpi_cost_minimised} is minimal for decreasing $N$, $ N >= 1,$ \: $N  \in \mathbb{N}$. This is the motivation for using six blocking messages for halo exchanges in Ludwig.

	\paragraph{}
	Thus, although the non-blocking version sends and receives the same amount of data, the greater number of messages involved is a similarly weighting trade off to the limitations of blocking three times; or such is the subject of this investigation.

\begin{figure}[H]
		\centering
		\includegraphics[width=0.99\textwidth]{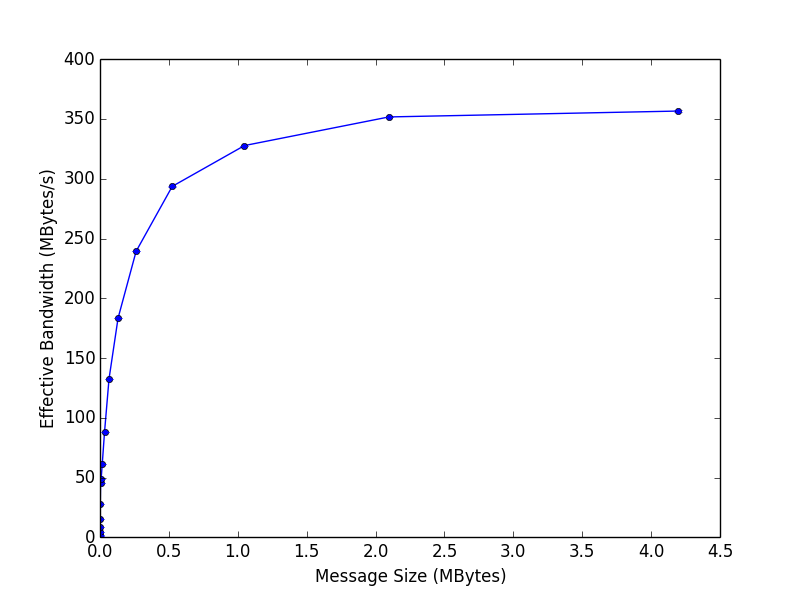}
		\caption{Bandwidth versus message size on ARCHER system.}
		\label{fig:ping_pong}
	\end{figure}

	\paragraph{}
	Figure \ref{fig:ping_pong} shows that the amount of data that is passed by MPI messages per second (Effective Bandwidth- see section \ref{sec:effective_bandwidth}) increases rapidly as we increase the size of messages sent and quickly reaches a plateau. At this point it is said that the Bandwidth has been saturated\footnote{A saturated Bandwidth is when messages are large enough to make use of the full Bandwidth of the network i.e. $m >= B$.}. This data was generated using a ping-pong test\footnote{A program that implements a back-and-forth message exchange between two MPI Tasks in order to determine the cost of message exchanges between them.} on ARCHER (the UK national supercomputer- see section \ref{sec:hardware}). It gives an insight to the network efficiencies of the system.
	
	\paragraph{}
	It is desirable to maximise Effective Bandwidth so a programer will generally aim to have a message size large enough that it just about saturates the systems Bandwidth.
	
	\paragraph{}
	It can be seen that the Bandwidth is mostly saturated with a message size of approximately $0.5 MBytes$. Maximum Effective Bandwidth for the system is approximately $350 MBytes/s$.
		
\chapter{Code Design} \label{chapter:development}

    \section{Conditions of Development} \label{sec:conditions_of_development}
    
    \paragraph{}
    The aim of the code development process was to develop a non-blocking halo exchange routine. The following aims were taken into careful consideration during the development process.
    
    \begin{itemize}
    
        \item The new routine should perform halo swaps in a non-blocking way; sending 26 messages per process.
        
	    \item The non-blocking version should result in the exact same data transfer between MPI tasks and thus give the same simulation results in all cases.
	    
	    \item The non-blocking routine should be split into two mains functions: a first, which starts the non-blocking communication; a second, which ends it.
	    
	    \item The final non-blocking version should not significantly reduce the performance of the code in any situation when compared to the blocking-via-copy version \\(\texttt{lb\_halo\_via\_copy}, see section \ref{sec:blocking}). Furthermore, it is desired to see an improvement by the non-blocking version, especially when work and communication is overlapped.
	    
	    \item The overall structure of the code should not be changed and it should be easy to switch between versions.
	    
    \end{itemize}
    
    \section{MPI\_Issend and MPI\_Irecv} \label{sends_recvs}
    
    \paragraph{}
    The Non-Blocking, and Blocking versions of the code referred to throughout the report are the halo exchange routines outlined in sections \ref{sec:non-blocking} and \ref{sec:blocking}  respectively. These terms do not refer to the types of MPI messages used for communication. Both of these versions use non-blocking messages: \texttt{MPI\_Issend} and \texttt{MPI\_Irecv}, to send and receive messages.
    
    \paragraph{MPI\_recv}
    The non-blocking receive is posted as soon as possible in each version of the code since the message can't arrive without it. An \texttt{MPI\_Waitall}, or an \texttt{MPI\_Waitany}, function acts as a barrier (the blocking point) until the relevant messages have completed.
       
    \paragraph{MPI\_Issend}
    In analogy to \texttt{MPI\_Isend}, \texttt{MPI\_Issend} is a synchronous non-blocking send. This means that the routine is thread-safe. As such it may be safely used by multiple threads without the need for any user-provided thread locks \cite{mpich}. Thread safety is not directly relevant in this report as threaded programming does not require halo exchanges.

    \section{Structure of the Code} \label{sec:structure_of_code}

    \paragraph{}
    At the beginning of each iteration of the simulation the \texttt{lb\_halo} function is called to swap the distributions at the periodic/processor boundaries in each direction. This calls either of the functions \texttt{lb\_halo\_via\_struct} or \texttt{lb\_halo\_via\_copy} depending on what type of model is being simulated.
        
    \paragraph{}
    \texttt{lb\_halo\_via\_struct} uses MPI datatypes to perform distribution halo swaps. It has a performance advantage over \texttt{lb\_halo\_via\_copy} which copies halo values into arrays which are dynamically allocated to the size of the messages to be sent. \texttt{lb\_halo\_via\_copy} works for all models and is therefore used when \texttt{lb\_halo\_via\_struct} would not be a successful routine. Both routines send data in the same blocking fashion as described in section \ref{sec:blocking_halo_exchange}.

    \paragraph{}
    All benchmarks in this report use flat buffers to copy relevant data rather than MPI datatypes. This allows an accurate comparison of benchmarks to be conducted under fair conditions and for the results to be easily applied to all lattice simulation models. The halo exchange routines are located in the \texttt{model.c} source file.
        
        \subsection{Blocking} \label{sec:blocking}
        
        \paragraph{}
        For the purpose of this report the only function in \texttt{lb\_halo}, for blocking version of the code, is \texttt{lb\_halo\_via\_copy}.
        
        \paragraph{\texttt{lb\_halo\_via\_copy}}
        For each dimension, this function: copies relevant data to buffers; posts receives; post sends and finally blocks, using the \texttt{MPI\_Waitall} function. This results in three blocks before communication is complete.

        \subsection{Non-Blocking} \label{sec:non-blocking}
             
        \paragraph{}
        Two new functions were introduced to be called instead of \texttt{lb\_halo}:
        
        \begin{enumerate}
	        \item \texttt{lb\_halo\_via\_copy\_nonblocking\_start()}
	        \item \texttt{lb\_halo\_via\_copy\_nonblocking\_end()}
        \end{enumerate}
        
        \paragraph{}
        These combine to perform the halo exchange routine. This routine is analogy to the \\
        \texttt{lb\_halo\_via\_copy} routine as it also uses a flat buffer to copy relevant data rather than MPI datatypes.
        
        \paragraph{}
        The non blocking routine is split into a \textit{start} and an \textit{end} function so that further work may be included between these two functions. In this way, work and communication can overlap, see section \ref{sec:overlap_comm_work}. 
        
        \paragraph{\texttt{lb\_halo\_via\_copy\_nonblocking\_start()}}
        This contains three functions: 
        
        \begin{enumerate}
	        \item \texttt{halo\_planes()}
	        \item \texttt{halo\_edges()}
	        \item \texttt{halo\_corners()}
        \end{enumerate}
        
        which each perform the following tasks in the respective order: post non-blocking receives; copy relevant data to be sent into buffers and finally, post the non-blocking sends. The sends are posted as soon as the relevant buffers have been filled. With this, the MPI communication has begun. There is no barrier (block) in this function.
        
        \paragraph{\texttt{lb\_halo\_via\_copy\_nonblocking\_end()}}
        When this function is called the program blocks until all receives have completed:
        
        \begin{lstlisting}
for (n=0; n<26; n++){
	MPI_Waitany(26, lb->hl.recvreq, &recvcount, lb->hl.status);
}
        \end{lstlisting}
        
        the receive buffers are then unpacked using \texttt{unpack\_halo\_buffers()}. The function is finally completed with another \texttt{MPI\_Waitany} for the sends.
        
        \paragraph{}
        Since the halo exchange routine is split into a \texttt{start} and \texttt{end} function, it is not possible to use locally declared arrays for the send and receive messages. Instead, a struct, \texttt{halo\_s} is declared, with variables for sending messages back and forth in each dimension. The size of these arrays are declared dynamically\footnote{At runtime.} within the \texttt{start} function. This struct is located in \texttt{lb\_model\_s.h}.
        
    \section{MPI Neighbour Ranks} \label{sec:neighbour_ranks} 
    
    \paragraph{}
    Each MPI Task must know the ranks\footnote{A number which acts as an identifier for the MPI tasks.} of the neighbours with which it will communicate. The neighbour's ranks, for each process, are determined at the beginning of the simulation and then saved in an array to be referenced when required. Any sends/receives posted with incorrect destination/source ranks will stall the program indefinitely. Code in this section is located in the \texttt{coords.c} source file.
    
    \paragraph{}
    A virtual topology is created for the ranks of the MPI Tasks which corresponds directly to data locality within the simulation.
    		  
    \begin{lstlisting}
MPI_Cart_create(pe_comm(), 3, pe_cartesian_size, iperiodic, reorder_, &cartesian_communicator);
    \end{lstlisting} 
    
    creates the virtual topology communicator (\texttt{cartesian\_communicator}) out of the processes in \texttt{pe\_comm()} communicator with 3 dimensions, each of size \texttt{pe\_cartesian\_size}. \texttt{iperiodic} is a 3 dimensional array with boolean values that determines whether or not to use periodic boundary conditions in each of these directions. \texttt{reorder\_} determines if the rankings may be reordered.
    		 
    	\subsection{Blocking} \label{sec:neighbr_ranks_block}
		
		\paragraph{}
		As discussed in section \ref{sec:blocking}, each process must communicate with a total of six neighbours- the ones orthogonally adjacent to it. The ranks of these are stored in an array of six integers with dimensions $2 \times 3$: \texttt{pe\_cartesian\_neighbours[dir][dim]}, where $\texttt{dir} = 2$ and $\texttt{dim} = 3$.
		
		\paragraph{}
		The values of \texttt{pe\_cartesian\_neighbours} are set at the beginning of the simulation using 
		
   		\begin{lstlisting}
for (n = 0; n < 3; n++) {
  MPI_Cart_shift(cartesian_communicator, n, 1,
		 pe_cartesian_neighbours[BACKWARD] + n,
		 pe_cartesian_neighbours[FORWARD] + n);
}
    	\end{lstlisting}
    	
    	where $n$ is the direction (X, Y or Z).
		
		\paragraph{}
		The ranks of the neighbour processes are written into a $2 \times 3$ dimensional array in the interest of maintainability and readability. These are accessed with the arguments of \texttt{dir} equal to either FORWARD or BACKWARD and \texttt{dim} equal to either X, Y or Z. Where these are equivalent to the values 1, 0, 0, 1, 2 respectively.

     	\subsection{Non-Blocking} \label{sec:neighbr_ranks_nonblock}

		\paragraph{}
		As discussed in section \ref{sec:non-blocking}, each process must communicate with a total of 26 neighbours. These neighbours, in terms of cartesian coordinates relative to a given process, are at position $(x, y, z)$, where $x$ $y$ and $z$ take on each value: $-1$, $0$ and $1$. There are 27 permutations but $(0, 0, 0)$ is excluded as this is the coordinate of the process in question.
		
		\paragraph{}
		A 1-dimensional array, \texttt{nonblocking\_cartesian\_neighbours}, of size 26 is used to store the ranks of these neighbours. In the interest of maintainability and readability these are accessed using enums:

   		\begin{lstlisting}
enum Halo_neighbour{NNN = 0, NNM, NNP, NMN, NMM, NMP, NPN, NPM, NPP,
                    MNN, MNM, MNP, MMN, MMP, MPN, MPM, MPP,
                    PNN, PNM, PNP, PMN, PMM, PMP, PPN, PPM, PPP} neigh;
    	\end{lstlisting}

		\paragraph{}
		The reader and developer should view these three letters as XYZ on a cartesian grid, where N (Negative) refers to being displaced by $-1$ in that dimension; M (Middle) has a displacement of $0$ in that dimension and finally P (Positive) has a displacement of $+1$ in that dimension.
		
		\paragraph{}
		For example, 'MNP' to corresponds the position in the array that contains the value of the rank with the cartesian coordinates $(0, -1, +1)$ in the virtual topology relative to the process running.
		
\begin{figure}[H]
			\centering
			\includegraphics[width=0.99\textwidth]{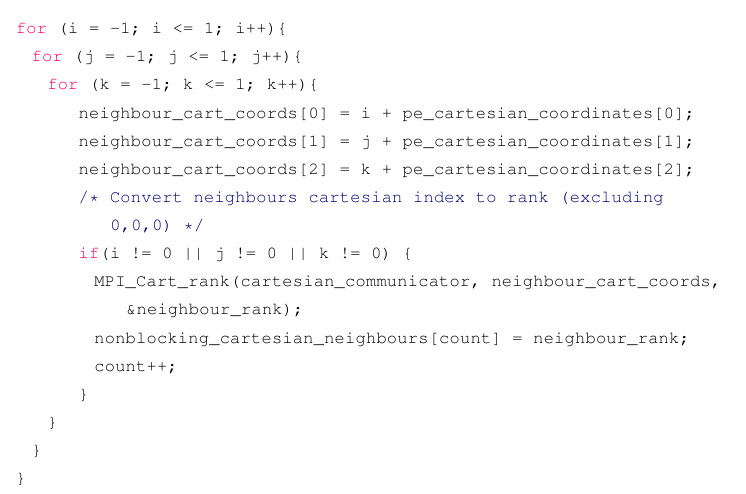}
			\caption{Code to fill the array \texttt{nonblocking\_cartesian\_neighbours} with the ranks of all 26 neighbours.}
			\label{fig:code}
		\end{figure}

		\paragraph{}
		The values of \texttt{nonblocking\_cartesian\_neighbours} are set at the beginning of the simulation using the code shown in figure \ref{fig:code}.
		
   		Consider the above code snippet. The three for loops go over all the permutations of $(x, y, z)$ with the values -1, 0, 1 to find the cartesian neighbours. For each permutation it sets 

   		\begin{lstlisting}
int neighbour_cart_coords[3],
    	\end{lstlisting}
    	
    	to these cartesian values. For each of these permutations, excluding $(0, 0, 0)$, 
    	
   		\begin{lstlisting}
MPI_Cart_rank(cartesian_communicator, neighbour_cart_coords, &neighbour_rank)
    	\end{lstlisting} 
    	
    	is used to determine the rank of the process located at those cartesian coordinates. Finally, the ranks fill the array
    	 
   		\begin{lstlisting}
int nonblocking_cartesian_neighbours[26]
    	\end{lstlisting}
    	
    	\paragraph{}
    	Note that the ordering of the for loops is essential to ensure it corresponds to the declared enumeration values discussed.

	\section{Overlapping Work and Communication} \label{sec:overlap_comm_work}
	
	\paragraph{}
	To see the potential advantages of the non-blocking version it is important to show the potential work-communication overlap. See chapter \ref{chapter:overlap}. To include work between the \texttt{start} and \texttt{end} routines the work must not be dependent on the halo data. To implement this overlap the \texttt{ludwig.c} source file was edited to call the \texttt{build\_update\_map} function between the \texttt{start} and \texttt{end} functions:

   	\begin{lstlisting}
lb_halo_via_copy_nonblocking_start(ludwig->lb);
	
build_update_map(ludwig->collinfo, ludwig->map);
	
lb_halo_via_copy_nonblocking_end(ludwig->lb);.
    \end{lstlisting}
	
		\subsection{build\_update\_map}
		
		\paragraph{}
		This function is relevant for simulations with solid particles in a fluid. The microscopic velocities in the fluid must not propagate to lattice sites within a solid object. Thus the simulation must map the location, moreover the location of the boundaries, of all solid objects each iteration.
		
		\paragraph{}
		This is independent of halo data and thus can be executed by idle processes while waiting for the completion of the communication of halo data on other processes.

	\section{Tests} \label{sec:tests}
	
	\paragraph{}
	As discussed in section \ref{sec:conditions_of_development}, the non-blocking version must give the exact same results as the blocking version for the simulation. Moreover, the exact same data must be transferred during halo exchange.
	
	\paragraph{}
	There were two main methods of testing used during the development process. These were regression tests and the unit tests. 
	
	\paragraph{Regression Tests} 
	There are ten standard regression tests provided with Ludwig. They check the output of ten simulations each ran on 8 MPI processes against a known to be correct set of outputs for the same set of simulations. Each simulation had a different input which were designed to test the code with very different conditions imposed. If the output was the same within a small degree of tolerence ($10^{-12}$) the test was passed. 
	
	\paragraph{\texttt{test\_halo()}}
	There are multiple unit tests for Ludwig. Each tests that a particular part, or unit, of the code is individually and independently scrutinized for proper operation. For the purpose of development during this project, only the halo exchange unit is relevant. \texttt{test\_halo} worked by setting the entire lattice on each subdomain to zero with the exception of the boundary values which are each set to different non-zero values. The halo exchange routine is then performed, when completed the halo values are explicitly compared with the boundary data on the appropriate neighbouring subdomains.
	
	\paragraph{}
	Section \ref{sec:development_process} will describe the way in which these tests were used to ensure a safe development process, and correctly implemented code.

	\section{Development Process} \label{sec:development_process}
	
	\paragraph{}
	The first stage of the development process was to develop how neighbours would be accessed in the non-blocking version. Once the new way to access neighbours, see figure \ref{fig:code}, had been developed, it was tested by using it in the blocking version. Any non-matching ranks in sends and receives would have resulted in the program stalling. Once the regression tests had all passed it was then possible to start development process of the non-blocking function. Not all neighbour rankings could be verified by the blocking version as it only communicates with six of the twenty-six neighbours, though it was indicated that the method of accessing the ranks of a neighbours was consistent. Ultimate verification of correct neighbour ranking was concluded as development of the non-blocking version proceeded as shall be discussed.
	
	\paragraph{}
	First the non-blocking code was developed in one function which started and ended the communication. Both the blocking version and the 'partly developed' non-blocking version were called during the regression tests. The blocking version was called first to fill the halo buffers with all the correct data. Then the nonblocking version re-wrote the halo data. In this way the code could be developed by testing the correctness of individual messages sent by the non-blocking version. This also conclusively verified correct neighbour ranking. More generally the correctness of sending the planes, but not the edges or corners, could be verified as the regression tests would fail if the planes' halo data was being over-written by incorrect values. This implementation process was followed for all three functions discussed in section \ref{sec:non-blocking}.
	
	\paragraph{}
	Since the $D3Q19$ model was used throughout development, as well as benchmarking, the regression tests will pass regardless of whether or not the corner halo data was being transferred correctly as the corners are not required for this model, see figure \ref{fig:D3Q19}. Thus this function was developed last and the \texttt{test\_halo()} unit test could thereby be used to test the whole halo swapping routine and therein verify the correctness of the \texttt{halo\_corners()} function.

\chapter{Methodology and Conditions of Simulations} \label{chapter:methodology}

	\section{Benchmark Methodology}
	
	\paragraph{}
	The main aim to of analysis in this report is to compare the efficiencies of the blocking and non-blocking versions of the code. Any significant change in performance of the two versions appear in the timing of the halo exchanges. Thus, it is the time spent in halo communication which is used for analysis in each case, with the exception of the work-communication overlap benchmarks as explained in chapter \ref{chapter:overlap}.
	
	\paragraph{}
	The $D3Q19$ model is used for all benchmarks. As can be seen in figure \ref{fig:D3Q19}, this version does not require the corner lattice halos. Thus it would be possible to further improve performance in the non-blocking version for this particular model by excluding the eight messages that send/receive to/from the corners. However, it is desirable to obtain results that are consistent and reliable for all potential models by ensuring that every buffer that is filled by the blocking version is also filled by the non-blocking version.
	
	\paragraph{}
	The determination of statistical errors is of central importance in order to assess the quality of a measurement and predict the amount of resources needed for reaching a specific accuracy goal. Each benchmark was executed five times. An average of these values was taken as the accepted time. The standard deviation was used as an error bar for each point. The error bars in the graphs are mostly very small. In the cases that the dependent variable in the graph was a function of the independent variable then this function was calculated before taking the standard deviation. The Standard Deviation ($\sigma$) was calculated by 
	
	\begin{align}
		\sigma = \sqrt{\dfrac{1}{N} \sum_{i=1}^{N}( O_i - \langle O \rangle )^2},	\label{eq:standard_deviation}
	\end{align}
	
    where N is the number of measurements to be averaged (5 for benchmarks in this project), $O_i$ is the ith measurement of the given observable for the given benchmark and $\langle O \rangle = \dfrac{1}{N} \sum_{i=1}^{N} O_i$.
	
	\paragraph{}
	Decomposition is referred to as the way in which the dimensions of the overall lattice is decomposed into subdomains. A decomposition of $x \times y \times z$ is the number of subdomains aligned along the x y and z directions respectively. The number of MPI tasks (or subdomains) in the simulation is then easily retraced by compounding these values. Similarly, Local domain, or subdomain size, refers to the number of lattice sites in each dimension on the subdomains and follows the same structure: $x \times y \times z$.
	
	\paragraph{}
	Weak scaling is the process of scaling problem size with number of MPI tasks. The desired goal being to increase problem size without significantly increasing runtime by simply using more cores for the simulation. Strong scaling involves maintaining a constant problem size or constant number of MPI tasks  while varying the other. Strong scaling causes the problem size on each subdomain to change and will hence effect the ratio of communication to work on each subdomain, as well as the sizes of the messages being sent. Thus, strong scaling was used to uncover the differences between the blocking and non-blocking versions in this project.
	
	\paragraph{}
	Chapter \ref{chapter:1_node} focuses on determining how each version of the code performs with different subdomain sizes while maintaining a constant number of MPI tasks. The aim of this is to highlight the effect of message size and communication to work ratio on the relative performance of the non-blocking version to the blocking version. This is done for both cubic and non-cubic subdomains with the aim of isolating the effect of the the communication to work ratio as this is the only significant difference between cubic and non-cubic subdomains.

    \paragraph{}
    The most typical manner of assessing the performance of a parallel program  in respect of strong scaling is its speedup, the ratio of the speed of the simulation on $p$ processors (or equivalently, number of MPI tasks) as a multiple of its speed on basic (or minimum) number of cores. Speedup is given by
    
    \begin{align}
	    S(p) = \dfrac{T_1}{T_p},
	    \label{eq:speedup}
    \end{align}  
    
    where $T_1$ is the runtime on the minimum number of processes and $T_p$ is runtime on $p$ processes.
    
    \paragraph{}
    One might also choose to use parallel efficiency, defined as
    
    \begin{align}
	    E(p) = \dfrac{S(p)}{p}.
    \end{align}
    
    where $E(p) = 1$ is the same as an ideal linear speedup.
    
    \paragraph{}
    The focus of chapter \ref{chapter:strong_scaling} is determine how the performance of the non-blocking version, relative to the blocking version, scales onto a large number of cores. It is the aim of this chapter to determine if desynchronicity (non-blocking) is an advantage or disadvantage for a large number of MPI tasks. This is determined by examining the speedup and parallel efficiency of each version on a sensible number of cores ie. when parallel efficiency is above approximately 80\%.
    
    \section{Systematic Error} \label{sec:systematic_error}
    
    \paragraph{}
    There exists a source of systematic error in the way the benchmarks were conducted. When a job was submitted to the supercomputer it was specified to execute each benchmark five times. These jobs would then be completed consecutively within the same job. Thus using the same nodes and cores for each of the five executions. This may result in small inconsistencies where results are sometimes incorrect and not within standard error since the network used for each version has been different.
    
    \paragraph{}
    For the most part this was avoided by running all benchmarks in the same script when possible. This was a particularly notable necessity when the runtime of the non-blocking with overlapped work and communication had a slightly longer runtime than the non-blocking version without overlapped work and communication- outside the tolerance of standard error.
        
    \section{Hardware} \label{sec:hardware}
    
    \paragraph{ARCHER}
    is the latest UK National Supercomputing Service \cite{archer_guide}. The ARCHER hardware consists of the Cray XC30 MPP supercomputer containing external login nodes, postprocessing nodes and associated filesystems. Each postprocessing node contains 12-core Intel Ivy Bridge series processors which are dual processing cores. Thus, each processing node effectively contains 24 processing cores. There are 4,920 postprocessing nodes on archer, therefore there are $4,920 \times 24 = 118,080$ processing cores and potentially that many concurrent MPI tasks available for well scaling simulations.
    
    \paragraph{}
    Benchmarks of Ludwig on ARCHER in this report are taken using full computing nodes resulting in the number of MPI tasks always being divisible by 24. Each node is a shared memory environment and so using a varied number of cores per processing node may give inconsistent estimates of performance as nodes will have an unpredictable access to the node's memory bandwidth.

\chapter{Subdomain Size} \label{chapter:1_node}

	\section{Communication versus Work} \label{sec:communication_work}
	
	\paragraph{}
	Parallelism improves performance by decreasing subdomain lattice size and, identically, the number of sites that need to be updated by each core. Assuming all subdomains are cubic, work on each subdomain scales as 
	
	\begin{align}
		W \sim L^3,
		\label{eq:work}
	\end{align}
	
	where L is the length of each dimension in a cubic lattice.
	
	\paragraph{}
	Then communication \Big($\Big(\frac{\sum_{1}^{N}m}{\sum_{1}^{N}B}\Big)$ for fixed N \Big) scales as:
	
	\begin{align}
		M \sim 6L^2 + 12L + 8,
		\label{eq:communication}
	\end{align}
	
	or simply $M \sim 6L^2$, where the first term accounts for the six planes; the second for the twelve edges and the third for the eight corners as in figure \ref{fig:non_blocking}. The same equation can be derived from blocking communication trivially.

\begin{figure}[h]
		\centering
		\includegraphics[width=0.8\textwidth]{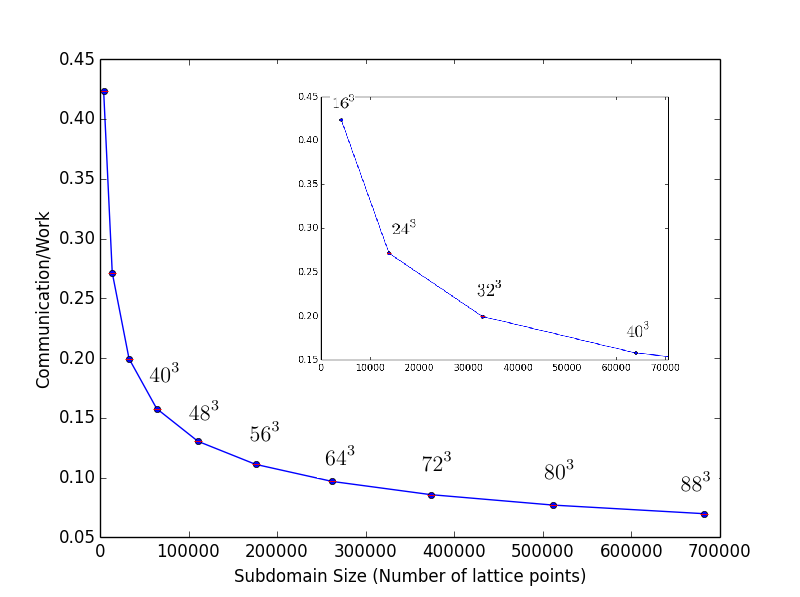}
		\caption{Ratio of communication to work for increasing subdomain size.}
		\label{fig:communicationVwork}
	\end{figure}

	\paragraph{}
	Figure \ref{fig:communicationVwork} shows that the overhead of communication to work decreases exponentially for increasing subdomain size. This ratio clearly becomes insignificant for large lattice sizes, although other factors such as core memory and cache sizes incur performance penalties as subdomain size is increased. Thus it is important to optimise performance on each core by determining the subdomain size with the lowest communication to work ratio before other aforementioned factors become significant.
	
	\section{Effective Bandwidth} \label{sec:effective_bandwidth}
	
	\paragraph{}
	The effect of local domain size on rate at which messages are sent can be examined by calculating the Effective Bandwidth for increasing subdomain lattice sizes. Effective Bandwidth ($B_{Eff}$) is defined here as the number of bytes sent per second (MBytes/second) as a function of the size of the messages being sent (in MBytes).
	
	\begin{align}
		B_{Eff} &= \dfrac{\text{(Number of halo lattice sites per subdomain)} \text{(Number of bytes per site)}}{\text{Time for one iteration}} \nonumber \\
		B_{Eff} &= \dfrac{(6L^2 + 12L + 8) \times 8 \times 19}{t}, \label{eq:effective_bandwidth}
	\end{align}
	
	where there are 8 bytes for each double and 19 doubles per lattice site. $t$ is time for one execution of the halo exchange routine. 
	
	\begin{figure}[H]
		\centering
		\begin{subfigure}[h]{0.85\textwidth}
			\centering
			\includegraphics[width=\textwidth]{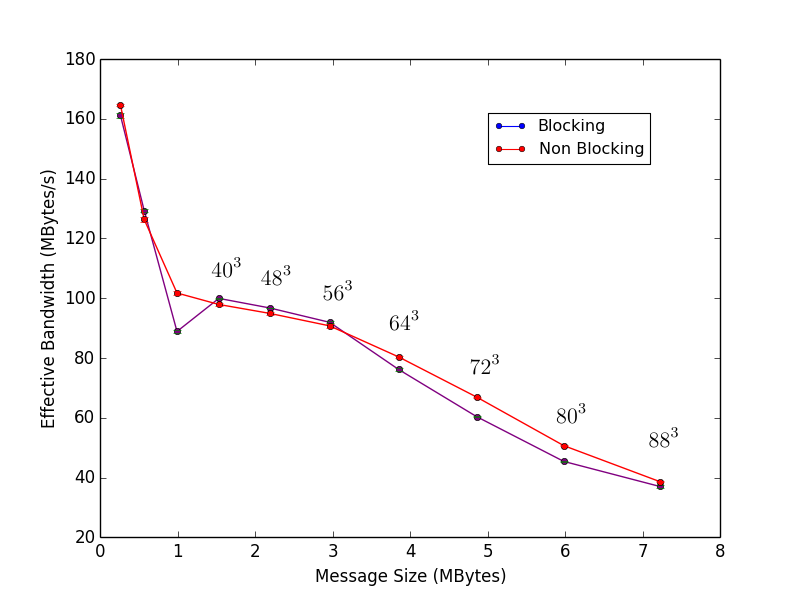}
			\caption{Subdomain lattice sizes ranging from $16^3$ to $88^3$.}
			\label{fig:1_node_EB.png}
		\end{subfigure}
		\hfill
		\begin{subfigure}[h]{0.85\textwidth}
			\centering
			\includegraphics[width=\textwidth]{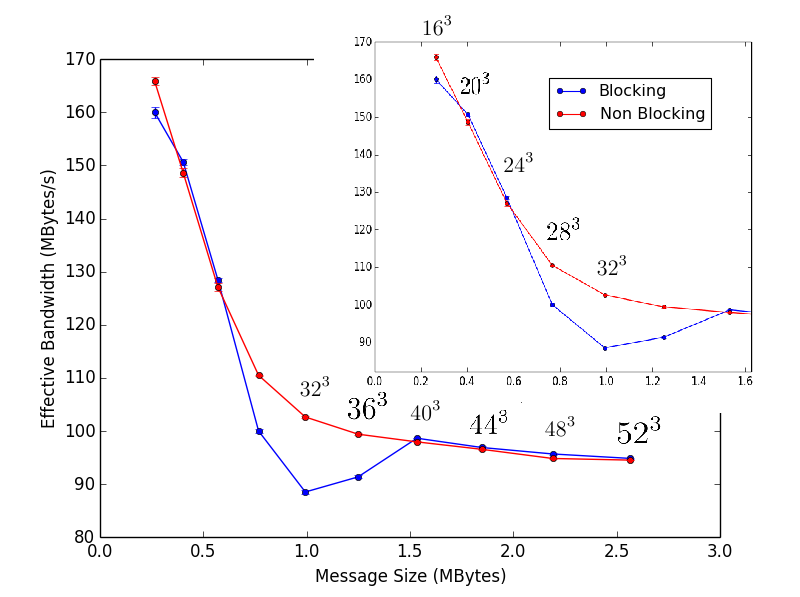}
			\caption{Subdomain lattice sizes ranging from $16^3$ to $52^3$.}
			\label{fig:1_node_EB_zoomed}
		\end{subfigure} 
		\caption{Effective Bandwidth. 1 node on ARCHER, 24 MPI tasks. Lattice decomposition $4 \times 3 \times 2$ with cubic subdomains as indicated.  Most error bars are smaller than points shown.}
		\label{fig:EB}
	\end{figure}

	\paragraph{}
	Figure \ref{fig:EB} shows the Effective Bandwidth for both the blocking and non-blocking versions of the code, calculated using equation \ref{eq:effective_bandwidth}. The number of MPI Tasks has been kept constant. The message size has been varied by changing the system size and therefore the subdomain size and the amount of halo data required.
	
	\paragraph{}
	Note a dip in performance for the blocking version of the code with local domain size of $32^3$. This is believed to be due to a change in protocol or standard buffer size which is hence inconsistent and omitted as irrelevant to general trend of the graph- see section \ref{sec:8_node_analysis} later.
	
	\paragraph{}
	Comparing figures \ref{fig:EB} and \ref{fig:ping_pong} it is possible to see that most of the message sizes used for these simulations with Ludwig have saturated the Bandwidth on ARCHER (see section \ref{sec:mpi_message_cost}). In analogy to the ping-pong test the Effective Bandwidth does not plateau but decreases as message size increases. This is because $t$ is the measure of the time taken for the whole halo exchange routine which includes the time taken in copying the halo data into and out of buffers. Thus the slight difference that can be seen in the Effective Bandwidth between these two versions is due to blocking time (for the blocking version), and the time to send extra messages (for the non-blocking version).
	
\begin{figure}[H]
		\centering
		\includegraphics[width=0.99\textwidth]{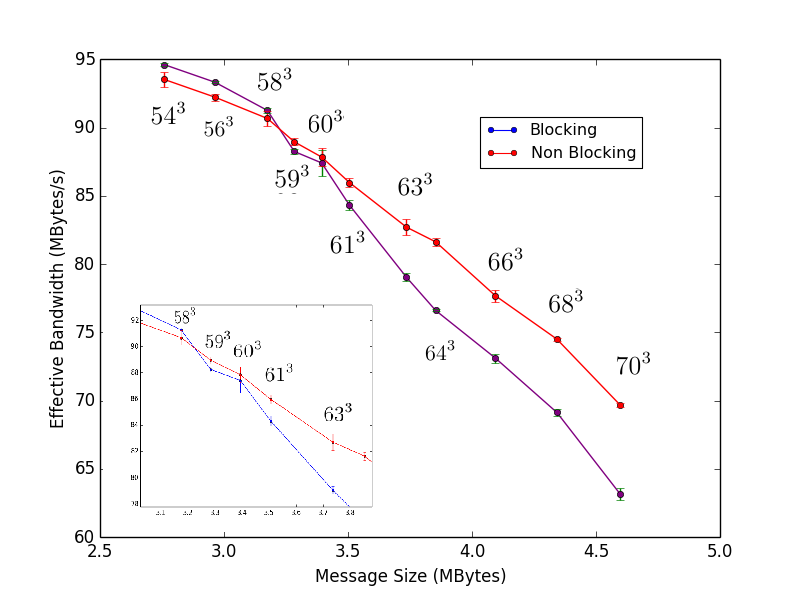}
		\caption{1 node on ARCHER, 24 MPI tasks. Lattice decomposition $4 \times 3 \times 2$ with cubic subdomains as indicated.  Most error bars are smaller than points shown.}
		\label{fig:1_node_EB_overtake}
	\end{figure}

	\paragraph{}
	There is a similar Effective Bandwidth for both versions for message sizes less than 3 MBytes. For larger messages, however, the non-blocking version has a higher Effective Bandwidth. This difference is more significant than the small fluctuations of the performance between the non-blocking and blocking version for smaller messages. This can be seen more closely in figure \ref{fig:1_node_EB_overtake}.

\begin{figure}[H]
		\centering
		\includegraphics[width=0.9\textwidth]{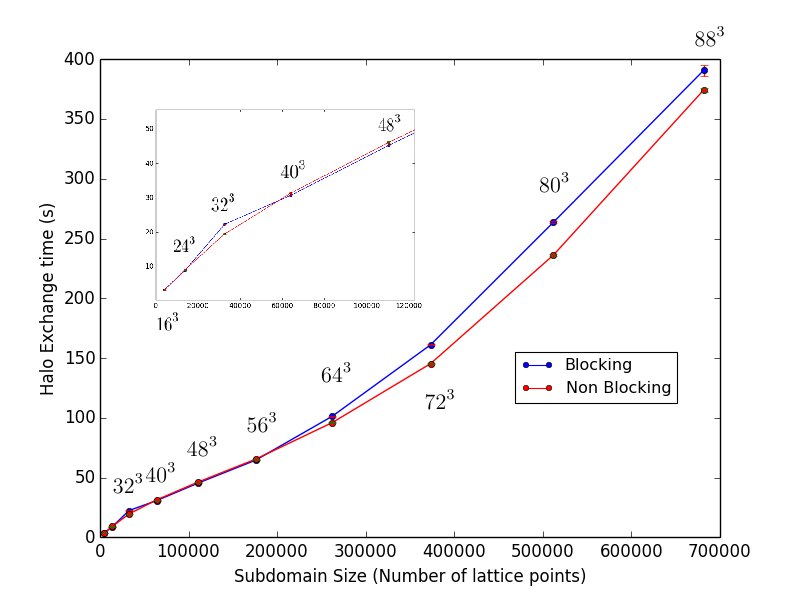}
		\caption{Time for $2,000$ calls to the lattice halos routine against subdomain size. 1 computing node on ARCHER; using 24 MPI Tasks. Lattice decomposition $4 \times 3 \times 2$ with cubic subdomains as indicated. Lattice sizes vary from $16^3$ to $88^3$.  Most error bars are smaller than points shown.}
		\label{fig:1_node_times}
	\end{figure}

	\paragraph{}
	The raw timings used to produce the Effective Bandwidth can be seen in figure \ref{fig:1_node_times}. The performance of the blocking code appeared to be slightly better for small subdomain sizes. However it is clear from figure \ref{fig:1_node_times} that the advantage that the non-blocking version has for large messages is considerably more significant.

	\paragraph{}
	The subdomain size that is usually chosen, in order to maximise the balance between runtime and parallel efficiency, is between $24^3$ and $48^3$. It is notable that there is only a small difference in performance of the blocking and non-blocking versions in this region.

	\section{Rate of Updates} \label{sec:updates_per_sec}
	
	\paragraph{}
	It is possible to determine the efficiencies of the different versions of the code by examining the Number of Updates per Core per Second as subdomain size is varied. Varying subdomain size is analogy to changing message size as they are both a function of $L$.
	
	\paragraph{}
	The Number of Updates per Core per Second can be given by
	
	\begin{align}
		N = \dfrac{L^3}{t},
		\label{eq:updates}
	\end{align}
	
	where t is once again the time for one execution of the halo exchange routine.

	\newpage
	\begin{figure}[H]
		\centering
		\begin{subfigure}[h]{0.85\textwidth}
			\centering
			\includegraphics[width=\textwidth]{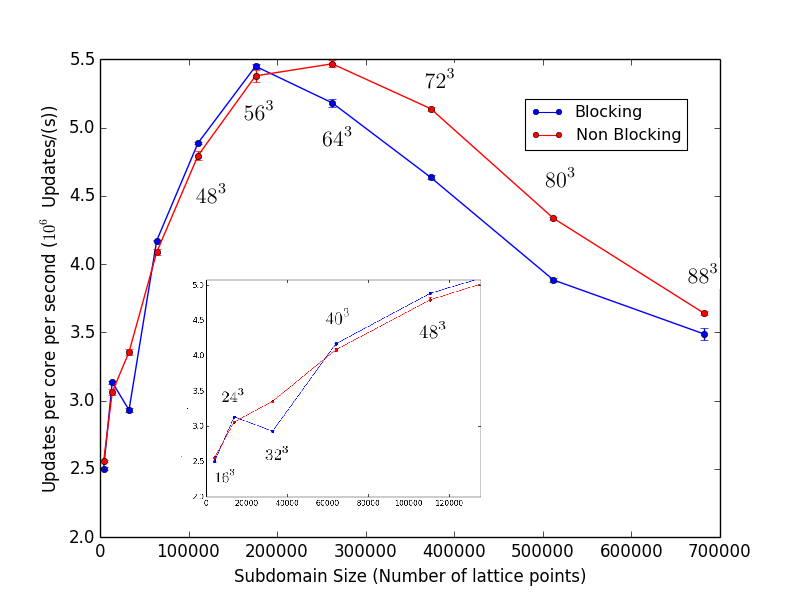}
			\caption{Subdomain lattice sizes ranging from $16^3$ to $88^3$.}
			\label{fig:1_node_updates_per_sec.png}
		\end{subfigure}
		\hfill
		\begin{subfigure}[h]{0.85\textwidth}
			\centering
			\includegraphics[width=\textwidth]{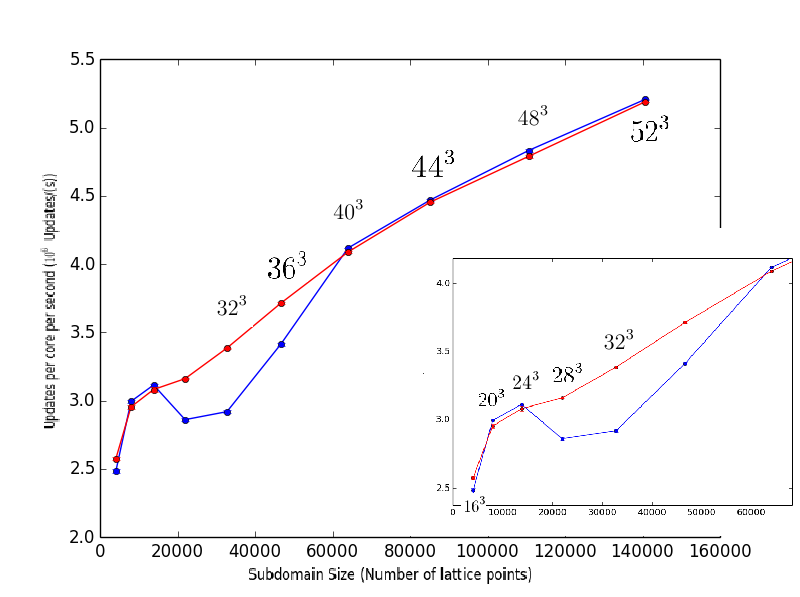}
			\caption{Subdomain lattice sizes ranging from $16^3$ to $52^3$.}
			\label{fig:1_node_updates_per_sec_zoomed.png}
		\end{subfigure} 
		\caption{1 node on ARCHER, 24 MPI tasks. Lattice decomposition $4 \times 3 \times 2$ with cubic subdomains as indicated. Updates per second ranges from $2.2 \times 10^6$ to $5.5 \times 10^6$.  Most error bars are smaller than points shown.}
		\label{fig:updates_per_sec}
	\end{figure}
	
 	\paragraph{}
 	Note, since t is not total iteration time, $N$ is actually the number of updates per core per second of time each site needs to spend in the halo exchange routine. Figure \ref{fig:updates_per_sec} shows $N$ for both the blocking and non-blocking versions of the code, calculated using equation \ref{eq:updates}.
	
	\paragraph{}
	Note another dip in performance for the blocking version of the code with local domain size of $32^3$. This is disregarded for the same reason as explained in section \ref{sec:effective_bandwidth}.
	
\begin{figure}[h]
		\centering
		\includegraphics[width=0.9\textwidth]{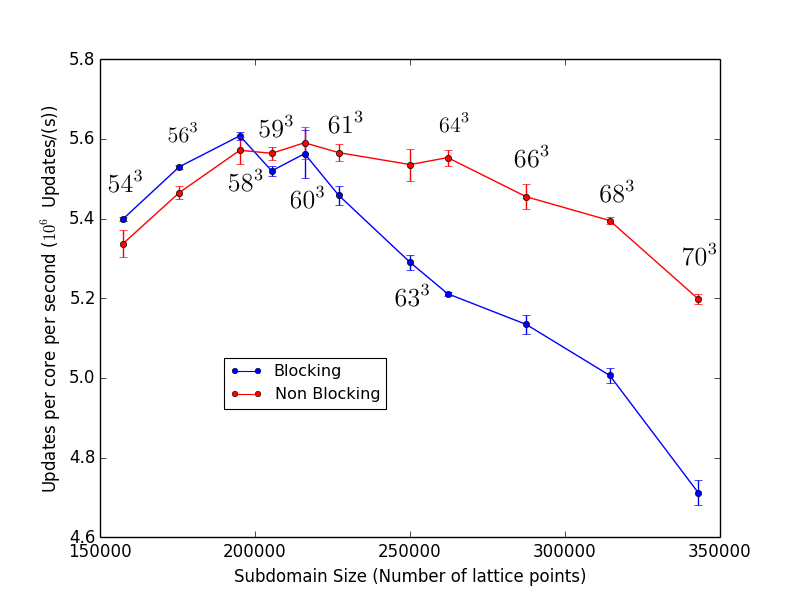}
		\caption{1 node on ARCHER, 24 MPI tasks. Lattice decomposition $4 \times 3 \times 2$ with cubic subdomains as indicated. Updates per second ranges from $2.2 \times 10^6$ to $5.5 \times 10^6$. }
		\label{fig:1_node_updates_per_sec_peak}
	\end{figure}

	\paragraph{}
	A turning point in the Number of Updates per Core per Second ($N$) can be seen between local domain sizes $48^3$ and $72^3$, emphasized in figure \ref{fig:1_node_updates_per_sec_peak}. This seems reasonable considering that significant reductions in the communication-work ratio plateaus at about 10\% in this region as seen in figure \ref{fig:communicationVwork}. The non-blocking version appears to reach its peak $N$ for a lattice size of $\approx 60^3 = 216,000$ rather than $\approx 58^3 = 195,112$ for the blcoking version, a difference of $20,888$ sites per subdomain. This suggests that the non-blocking version of the code may be a superior choice for simulations where we've chosen to use large subdomains for lack of core resources or otherwise.
	
	\paragraph{}
	It was discussed in section \ref{sec:mpi_message_cost}, and in particular equation \ref{eq:mpi_cost_minimised} that the advantage of the blocking version is that it minimises equation \ref{eq:mpi_cost_minimised} for a given simulation size. As subdomain dimension size ($L$) is increased $M$ also gets exponentially larger accordingly with equation \ref{eq:communication}. Thus, for large M, the second term of equation \ref{eq:mpi_cost_minimised} becomes the significant term and so there is a point at which the advantage of non-blocking becomes significant over the advantage of the reduced latency. 

	\paragraph{}
	Generally simulations will aim to have subdomain sizes in the range from $\approx 32^3$ to $\approx 48^3$. In this region there is no significant difference between the two versions.
	
	\section{Non Cubic Subdomain} \label{sec:non_cubic_subdomain}
	
		\subsection{Communication versus Work} \label{communicationVwork_noncubic}
		
		\paragraph{}
		In section \ref{sec:communication_work} an expression for how work (W) and communication (M) scale with lattice length L for a cubic lattice was derived. The ratio graphed in figure \ref{fig:communicationVwork} will clearly be different for a non-cubic lattice and there is a large number of combinations of subdomain dimensions to try. However, since Ludwig always strives for a sensible- with good load balance- decomposition, and since lattices with non-prime dimensions are certainly preferred, it is rare that any one dimension will be a vast multiple of another. Thus an insightful analysis of the performance with non cubic subdomains can be achieved by examining a trivial example with subdomain dimensions: $ x \times \frac{3}{2}x \times 2x$.
		
		Work may now scale as:
		
		\begin{align}
			W \sim& \quad (x) (1.5x) (2x) \nonumber \\
			W \sim& \quad 3x^3, \label{eq:work_noncubic}
		\end{align}
	
		where x is simply the number of lattice sites in the x-direction on the subdomain, chosen arbitrarily.
		
		\paragraph{}
		Communication now scales as
		
		\begin{align}
			M \sim& \quad 2(x^2) + 2(1.5x)^2 + 2(2x)^2 + 4x + 4(1.5x) + 4(2x) + 8 \nonumber \\
			M \sim& \quad 14.5x^2 + 18x + 8 \label{eq:communication_noncubic}
		\end{align}
		
		where the same logic as in section \ref{sec:communication_work} was followed.
		
\begin{figure}[H]
			\centering
			\includegraphics[width=0.9\textwidth]{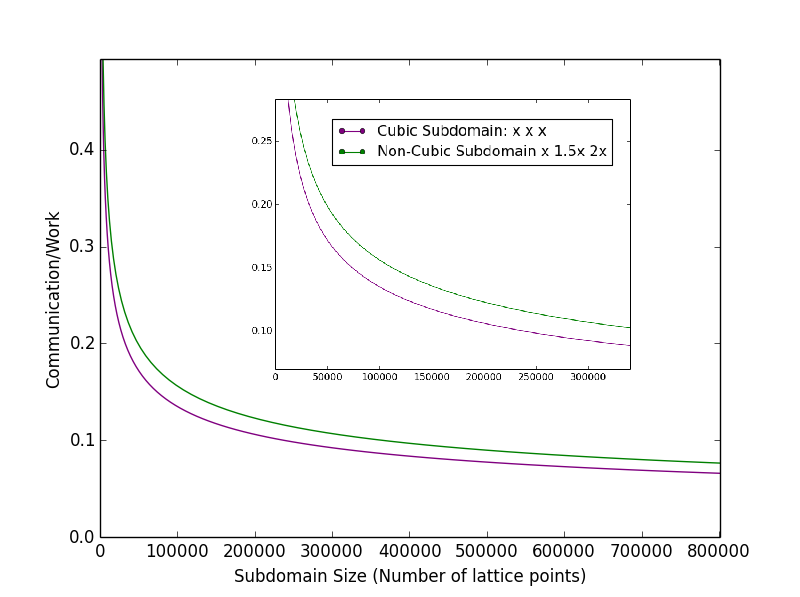}
			\caption{Comparison of the analytical ratio of communication to work for cubic and non-cubic subdomains as indicated.}
			\label{fig:communicationVwork_comparison}
		\end{figure}

		Figure \ref{fig:communicationVwork_comparison} shows an analytical solution to the ratio of communication to work for a cubic subdomain with dimensions: $x \times x \times x$ and a noncubic subdomain, with dimensions: $ x \times \frac{3}{2}x \times 2x$. 
		
		\subsection{Rate of Updates} \label{sec:effective_bandwidth_noncubic}
		
		\paragraph{}
		Figure \ref{fig:communicationVwork_comparison} demonstrates that simulations with less dimensionally balanced subdomains have a higher communication to work ratio. It has been shown in section \ref{sec:effective_bandwidth} that the non-blocking version has best performance relative to the blocking version for larger subdomains. Larger subdomains have larger messages (more communication), but a lower communication to work ratio. Thus the main difference in the analysis of non-cubic subdomains is determining the direct influence on the performance of the non-blocking version relative to blocking version due to the communication to work ratio. 
		
		\begin{figure}[H]
			\centering
			\includegraphics[width=0.9\textwidth]{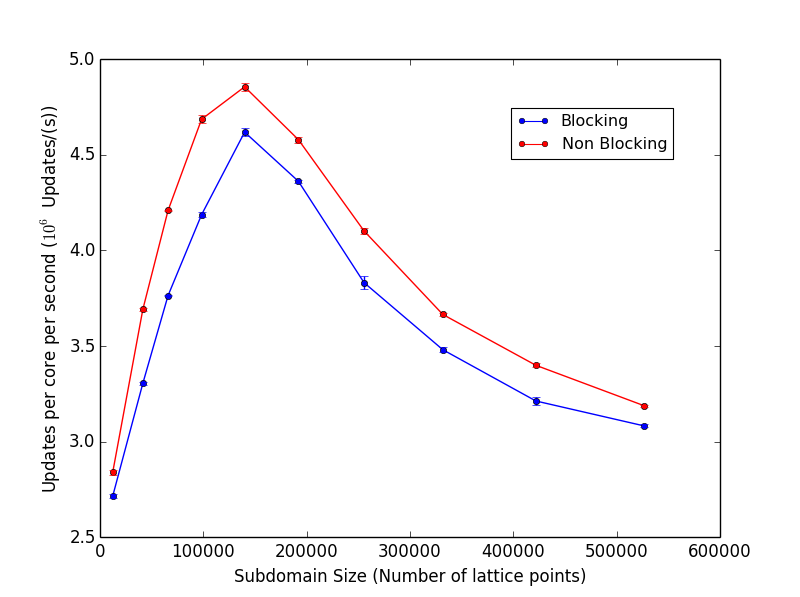}
			\caption{1 node on ARCHER, 24 MPI Tasks. Lattice decomposition $4 \times 3 \times 2$ with non-cubic subdomains of dimensions: $x \times \frac{3}{2}x \times 2x$. See table \ref{table:non-cubic} for details on lattice sizes.  Most error bars are smaller than points shown.}
			\label{fig:1_node_updates_per_sec_noncubic}
		\end{figure}
		
		\paragraph{}
		Consider figure \ref{fig:1_node_updates_per_sec_noncubic}. The number of updates per core $(N)$ for both versions is considerably less than for a cubic lattice when compared to figure \ref{fig:EB}, which is naturally the case due to the higher communication overhead. However, it is clear that there is a consistently higher $N$ for the non-blocking version of the code as opposed to the blocking version. This implies that the non-blocking code incurs a lower penalty when imposed with a higher communication-work ratio than its blocking counterpart.
		
		\paragraph{}
		Considering figure \ref{fig:1_node_updates_per_sec_noncubic} together with table \ref{table:non-cubic}, the peak in $N$ occurs at subdomain size of approximately $49^3$. The higher communication to ratio reduces both code versions efficiency for large subdomains.

		\begin{table}[H]
		\caption{Subdomain size and dimensions relating to figure \ref{fig:1_node_updates_per_sec_noncubic}.} 
		\centering 
			\begin{tabular}{c c c} 
			\hline\hline 
			Subdomain Dimensions & Subdomain Size & Size as cube \\ [0.5ex] 
			\hline 
			16 24 32 & 12,288 & $\approx 23^3$ \\
			24 36 48 & 41,472 & $\approx 35^3$ \\
			28 42 56 & 65,856 & $\approx 40^3$ \\
			32 48 64 & 98,304 & $\approx 46^3$ \\
			36 54 72 & 116,640 & $\approx 49^3$ \\
			40 60 80 & 192,000 & $\approx 58^3$ \\
			44 66 88 & 255,552 & $\approx 63^3$ \\
			48 72 96 & 331,776 & $\approx 69^3$ \\
			52 78 104 & 421,824 & $\approx 75^3$ \\
			56 84 112 & 526,848 & $\approx 81^3$ \\ [1ex] 
			\hline 
			\label{table:non-cubic}
			\end{tabular}
		\end{table}
		
	\section{8 Node Analysis} \label{sec:8_node_analysis}
	
	\paragraph{}
	Single node analysis may be considered a special case since all MPI communication is still within a shared memory environment. Thus this section is included to confirm trends seen throughout single node analysis in this chapter.
	
		\subsection{Effective Bandwidth and Updates per Core}
		
\begin{figure}[H]
			\centering
			\begin{figure}[H]
				\centering
				\includegraphics[width=0.75\textwidth]{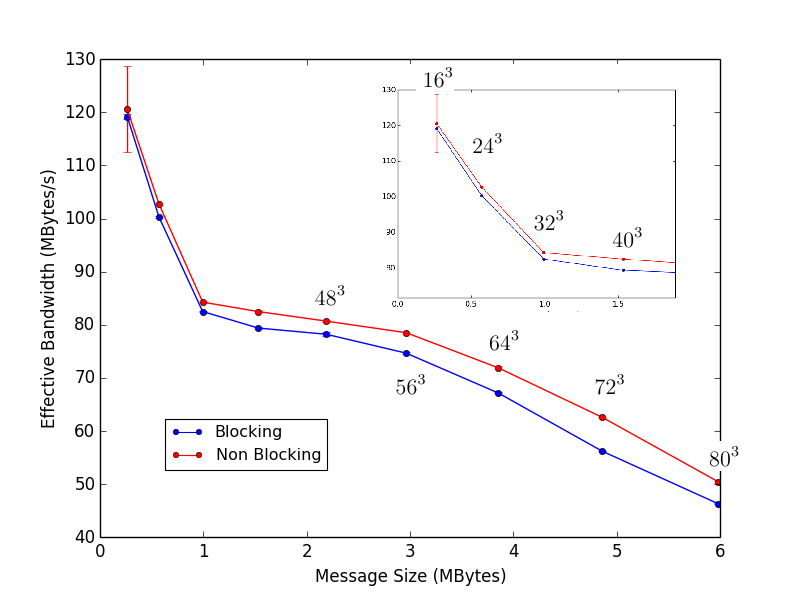}
				\caption{Effective Bandwidth.}
				\label{fig:8_nodes_EB}
			\end{figure}
			\begin{figure}[H]
				\centering
				\includegraphics[width=0.75\textwidth]{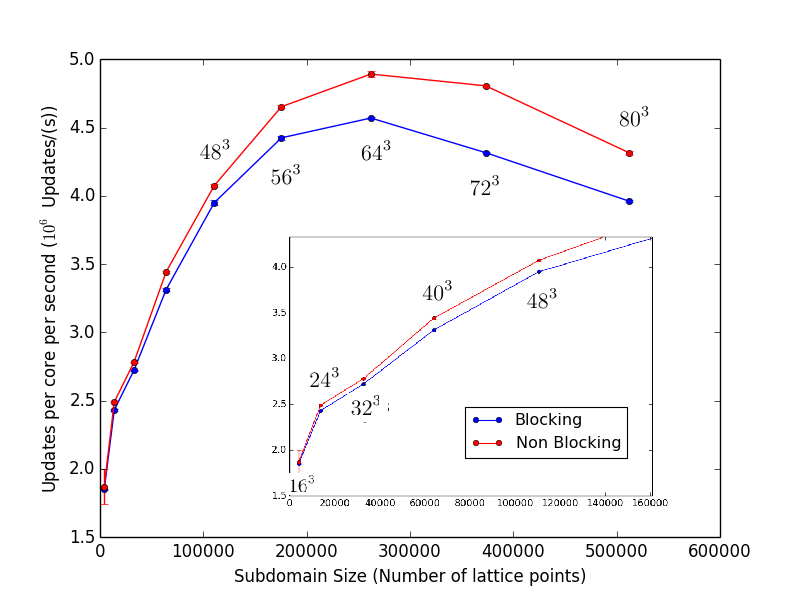}
				\caption{Updates per core $N$.}
				\label{fig:8_nodes_updates_per_core}
			\end{figure}
			\caption{8 nodes on ARCHER, 192 MPI Tasks. Lattice decomposition $8 \times 6 \times 4$ with cubic subdomains as indicated. Most error bars are smaller than points shown.}
			\label{fig:8_nodes}
		\end{figure}

		\paragraph{}
		The Effective Bandwidth ($B_{eff}$) and Number of Updates per Core per Second ($N$), using 8 full nodes on ARCHER (192 MPI Tasks) can be seen in figures \ref{fig:8_nodes_EB} and \ref{fig:8_nodes_updates_per_core} respectively. Comparing these to the $B_{eff}$ and $N$ in figures \ref{fig:1_node_EB.png} and \ref{fig:1_node_updates_per_sec.png}, a similar trend can be seen: both versions have a similar $B_{eff}$ and $N$ for small message sizes with non-blocking version gaining a greater advantage as the message size increases.
		
		\paragraph{}
		Since using 8 nodes ensures that the cores used do not all share the same memory environment, the trend is more pronounced, consistent and reliable for 8 nodes. This is reflected in graphs \ref{fig:8_nodes_EB} and \ref{fig:8_nodes_updates_per_core}.
		
\chapter{Strong Scaling Simulations} \label{chapter:strong_scaling}

\paragraph{}
In this chapter the effect of strong scaling on different lattice sizes is examined. The global lattice size is kept constant within each section and the number of MPI Tasks and therefore also the local domain size is varied. Chapter \ref{chapter:1_node} demonstrated that the non-blocking version of the code performs better on larger subdomain sizes and thus a good performance for the non-blocking version relative to the blocking version for a lower number of MPI tasks is expected. Though the main focus of this chapter is to examine the relative performance of the non-blocking to the blocking version for a large number of MPI tasks.

\paragraph{}
In this chapter Lattice size or system size refers to the number of lattice sites. Since the decomposition onto subdomains is dependent on the number of MPI Tasks the subdomains will generally not be cubic as they were in section \ref{sec:effective_bandwidth} and will have a difference decomposition for each simulation.

	\section{$96^3$ System Size} \label{sec:96_96_96}
	
\begin{figure}[H]
		\centering
		\includegraphics[width=0.99\textwidth]{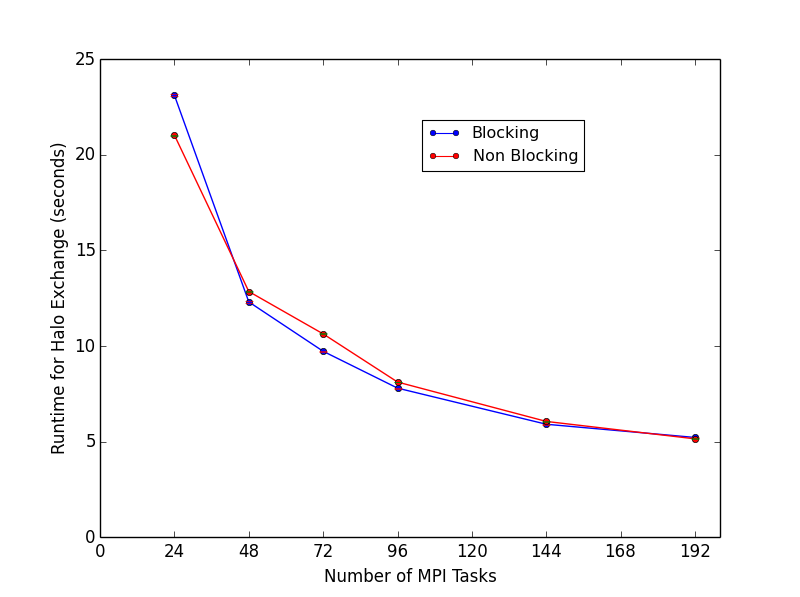}
		\caption{Time for 2,000 calls to halo exchange routine for system size $96^3$. Error Bars are small, mostly not visible under points.}
		\label{fig:96_seconds}
	\end{figure}

	\paragraph{}
	Figure \ref{fig:96_seconds} shows the execution time (in seconds) of the halo exchange function for increasing number of MPI Tasks. As the minimum number of MPI Tasks used is 24, the maximum subdomain size is still relatively small for all points here, as can be seen from table \ref{table:96_runtime}.

	\begin{table}[ht]
	\caption{Subdomain size and dimensions relating to figure \ref{fig:96_seconds}.} 
	\centering 
	\begin{tabular}{c c c} 
		\hline\hline 
		Number of Cores & Subdomain Dimensions & Subdomain Size \\ [0.5ex] 
		\hline 
		24 & 24 32 48 & 36,864 \\
		48 & 24 24 32 & 18,432 \\ 
		72 & 16 24 32 & 12,288 \\
		96 & 16 24 24 & 9,216 \\
		144 & 16 16 24 & 6,144 \\
		192 & 12 16 24 & 4,608 \\ [1ex] 
		\hline 
		\label{table:96_runtime}
		\end{tabular}
	\end{table}

		\subsection{Comparing Speedups} \label{sec:speedup_bad}
		
\begin{figure}[H]
			\centering
			\begin{subfigure}[h]{0.49\textwidth}
				\centering
				\includegraphics[width=\textwidth]{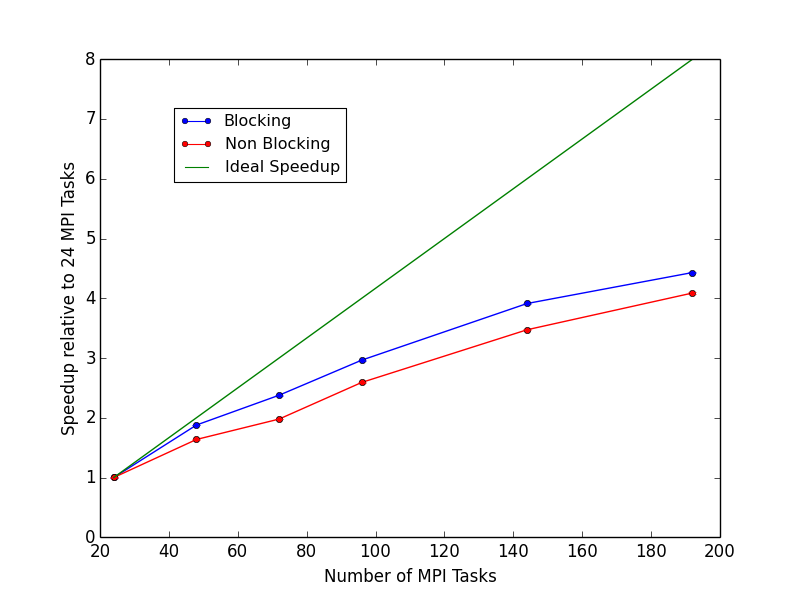}
				\caption{Speedup relative to respective $T_1$ values.}
				\label{fig:96_speedup}
			\end{subfigure}
			\hfill
			\begin{subfigure}[h]{0.49\textwidth}
				\centering
				\includegraphics[width=\textwidth]{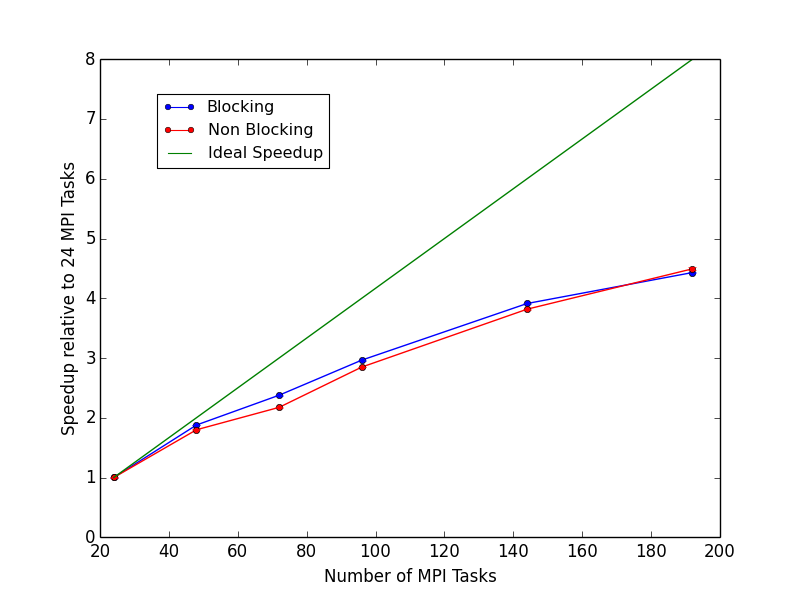}
				\caption{Speedup relative to same $T_1$ value.}
				\label{fig:96_speedup2}
			\end{subfigure} 
			\caption{Speedup for $96^3$ system size relative to 24 MPI Tasks. Error Bars are small, mostly not visible under points.}
			\label{fig:96_speedups}
		\end{figure}

		\paragraph{}
		It has been shown that the non-blocking version of the code has the best performance relative to the blocking version for large subdomain sizes. Thus, speedup may be biased toward the blocking version as the simulation with the largest subdomain size, which corresponds to the fewest MPI tasks (ie. $T_1$), will generally disproportionally favour the non-blocking version. This bias can be seen clearly in figure \ref{fig:96_speedup}, where the non-blocking version's speedup is systematically lower than the blocking version's. This is misleading when compared to the runtimes in figure \ref{fig:96_seconds}: the non-blocking version has a poorer speedup because it has a lower $T_1$ value. Although this is the correct speedup it is a poor comparison of the relative performance of each version on a large number of MPI tasks. Thus the speedup is replotted in figure \ref{fig:96_speedup2}, where the speedup has been calculated for each version relative to the same base runtime, $T_1$ (the runtime for 24 MPI tasks using the blocking version).  
		
		\paragraph{}
		Figure \ref{fig:96_speedup2} can now be considered as an accurate representation of how well the performance of each version scales onto a large number of cores. The speedups are very evenly balanced between versions. The parallel efficiency corresponding to figure \ref{fig:96_speedup2} is given in figure \ref{fig:96_efficiency2_appen} in the appendices.

	\section{$192^3$ System Size} \label{sec:192_192_192}

	\begin{figure}[H]
		\centering
		\includegraphics[width=0.99\textwidth]{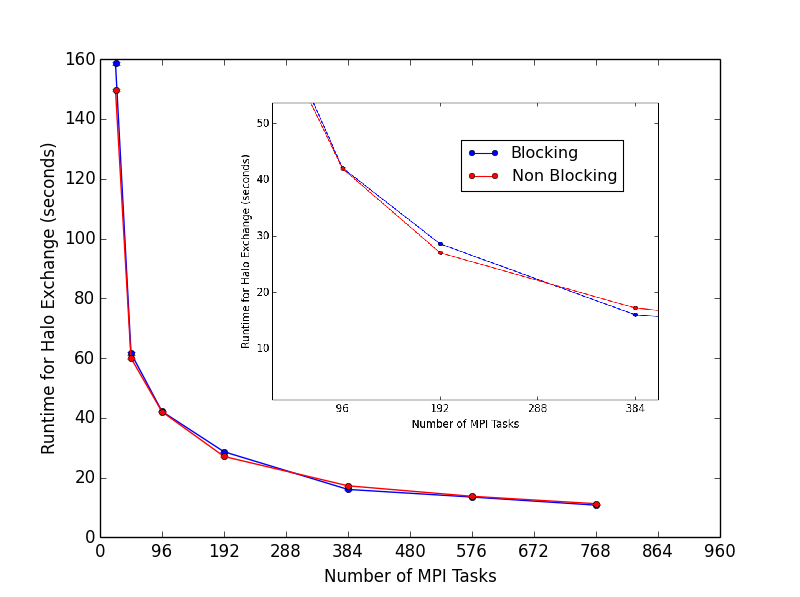}
		\caption{Time for 2,000 calls to halo exchange routine for system size of $192^3$. Error Bars are small, mostly not visible under points.}
		\label{fig:192_runtime}
	\end{figure}	

	\begin{table}[ht]
	\caption{Subdomain size and dimensions relating to figure \ref{fig:192_runtime}.} 
	\centering 
	\begin{tabular}{c c c} 
		\hline\hline 
		Number of Cores & Subdomain Dimensions & Subdomain Size \\ [0.5ex] 
		\hline 
		24 & 48 64 96 & 294,912 \\
		48 & 48 48 64 & 147,456 \\ 
		96 & 32 48 48 & 73,728 \\
		192 & 24 32 48 & 36,864 \\
		384 & 24 24 32 & 18,432 \\ 
		576 & 16 24 32 & 12,288 \\ 
		768 & 16 24 24 & 9,216 \\ [1ex]
		\hline 
		\label{table:192_runtime}
		\end{tabular}
	\end{table}
	
	\paragraph{}
	It is noticeable by considering figure \ref{fig:192_runtime} with table \ref{table:192_runtime} that there is a slightly faster runtime for the non-blocking version compared to the blocking version for less dimensionally balanced subdomains; especially when they are also larger subdomain sizes. This is noticeable for 24 cores and 192 cores.  
	
	\begin{figure}[H]
		\centering
		\begin{subfigure}[h]{0.49\textwidth}
			\centering
			\includegraphics[width=\textwidth]{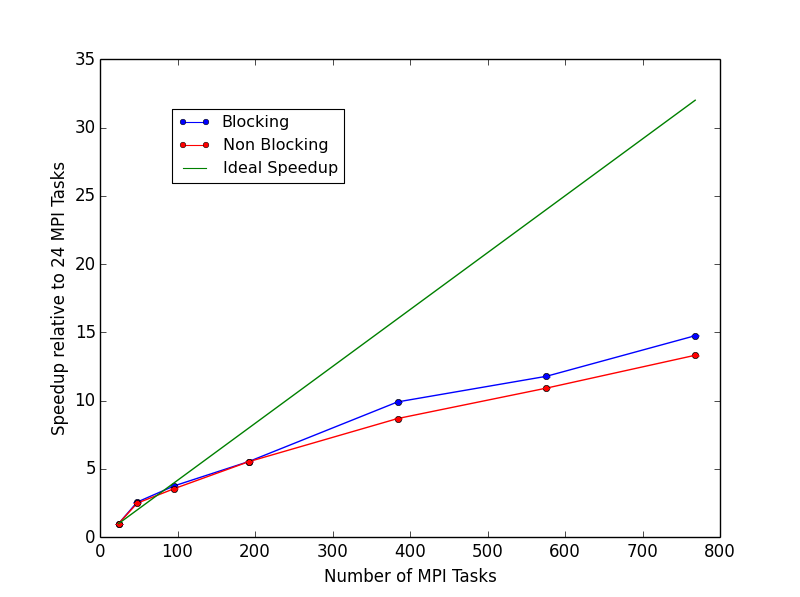}
			\caption{Speedup relative to respective $T_1$ values.}
			\label{fig:192_speedup}
		\end{subfigure}
		\hfill
		\begin{subfigure}[h]{0.49\textwidth}
			\centering
			\includegraphics[width=\textwidth]{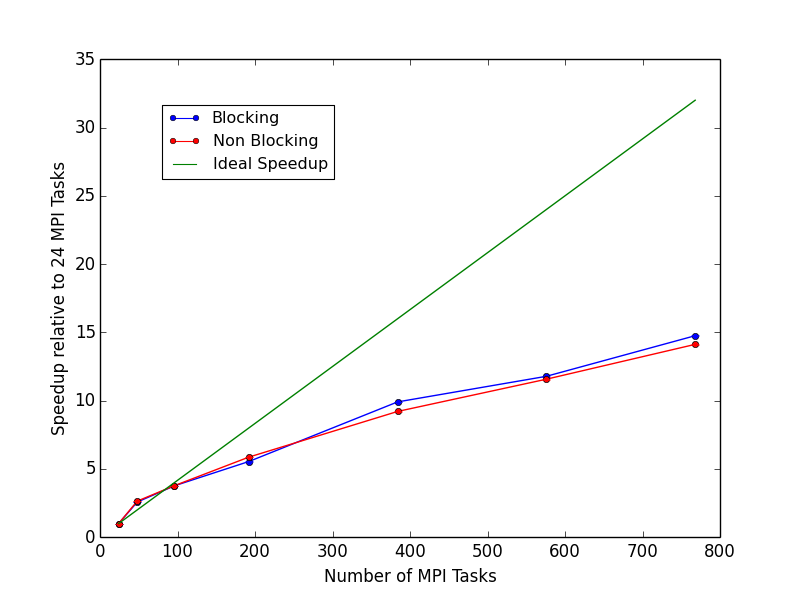}
			\caption{Speedup relative to same $T_1$ value.}
			\label{fig:192_speedup2}
		\end{subfigure} 
		\caption{Speedup for $192^3$ system size relative to 24 MPI Tasks. Error Bars are small, mostly not visible under points.}
		\label{fig:192_speedups}
	\end{figure}
	
	\paragraph{}
	From figure \ref{fig:192_speedup2} it can seen that the performance between both versions is similar as the number of MPI Tasks increases. The parallel efficiency corresponding to figure \ref{fig:192_speedup2} can be found in figure \ref{fig:192_efficiency2_appen} in the appendices.

	\section{$384^3$ System Size} \label{sec:384_384_384}
	
	\begin{figure}[H]
		\centering
		\includegraphics[width=0.99\textwidth]{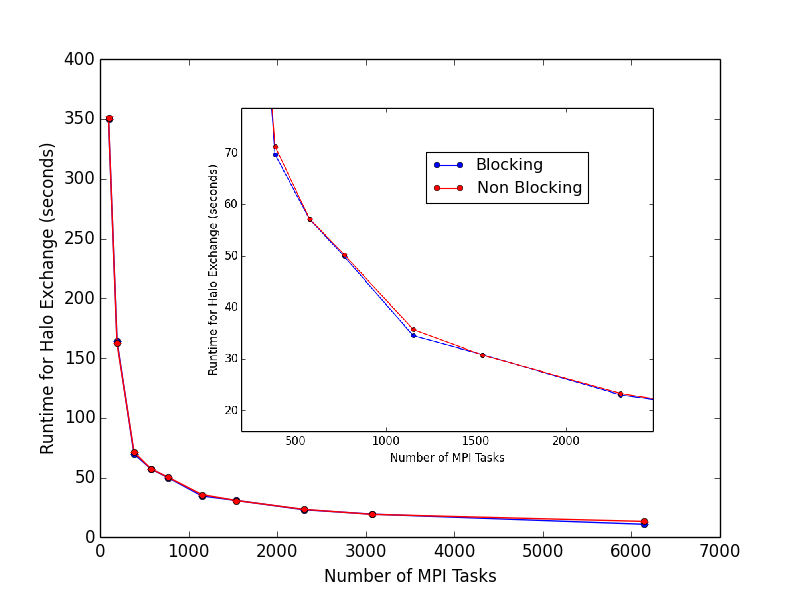}
		\caption{Time for 2,000 calls to halo exchange routine for system size $384^3$. Error Bars are small, mostly not visible under points.}
		\label{fig:384_runtime}
	\end{figure}

	\begin{table}[ht]
	\caption{Subdomain size and dimensions relating to figure \ref{fig:384_runtime}.} 
	\centering 
	\begin{tabular}{c c c} 
		\hline\hline 
		Number of Cores & Subdomain Dimensions & Subdomain Size \\ [0.5ex] 
		\hline 
		96 & 64 96 96 & 589,824 \\
		192 & 48 64 96 & 294,912 \\ 
		384 & 48 48 64 & 147,456 \\
		576 & 32 48 64 & 98,304 \\
		768 & 32 48 48 & 73,729 \\ 
		1,152 & 32 32 48 & 49,152 \\ 
		1,536 & 24 32 48 & 36,864 \\
		2,304 & 24 32 32 & 24,576 \\
		3,072 & 24 24 32 & 18,432 \\ 
		6,144 & 16 24 24 & 9,216 \\ [1ex]
		\hline 
		\label{table:384_runtime}
		\end{tabular}
	\end{table}
	
\begin{figure}[H]
		\centering
		\begin{subfigure}[h]{0.49\textwidth}
			\centering
			\includegraphics[width=\textwidth]{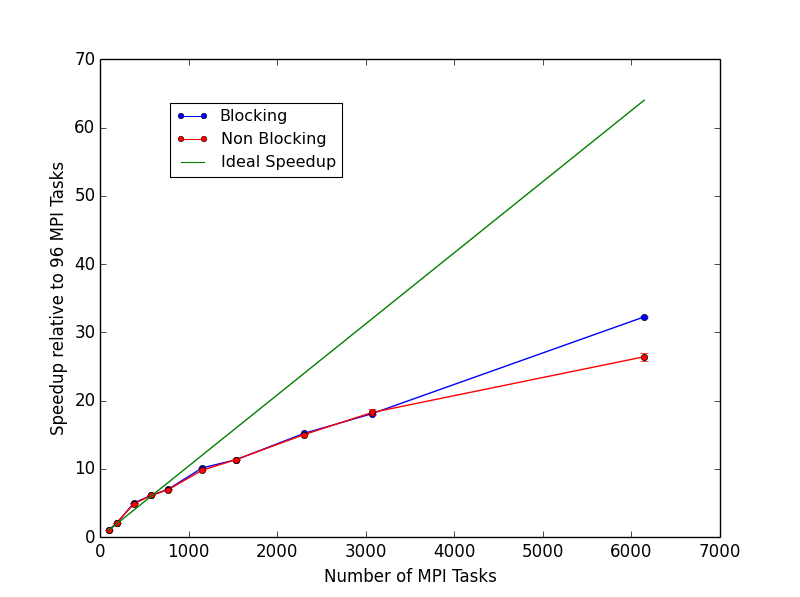}
			\caption{Speedup relative to respective $T_1$ values.}
			\label{fig:384_speedup}
		\end{subfigure}
		\hfill
		\begin{subfigure}[h]{0.49\textwidth}
			\centering
			\includegraphics[width=\textwidth]{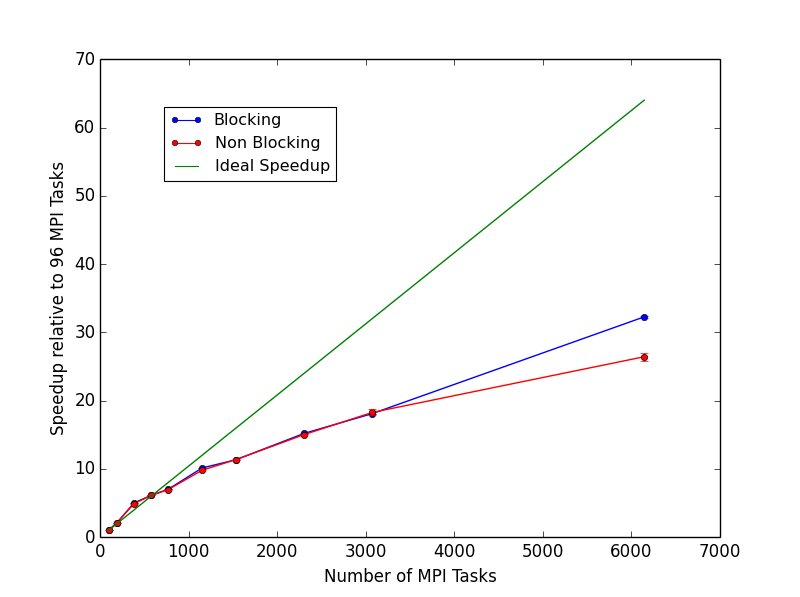}
			\caption{Speedup relative to same $T_1$ value.}
			\label{fig:384_speedup2}
		\end{subfigure} 
		\caption{Speedup for $384^3$ system size relative to 96 MPI Tasks. Error Bars are small, mostly not visible under points.}
		\label{fig:384_speedups}
	\end{figure}

	\paragraph{}
	Figure \ref{fig:384_runtime} shows the execution time of the halo exchange function for increasing number of MPI Tasks on a $384^3$ system size. The performance of both versions of the model are very evenly balanced, the runtimes are within standard error for most of the points. As before, both Speedup relative to each version's own $T_1$ and relative to the same $T_1$ are given in figure \ref{fig:384_speedups}.
	
\begin{figure}[H]
		\centering
		\includegraphics[width=0.99\textwidth]{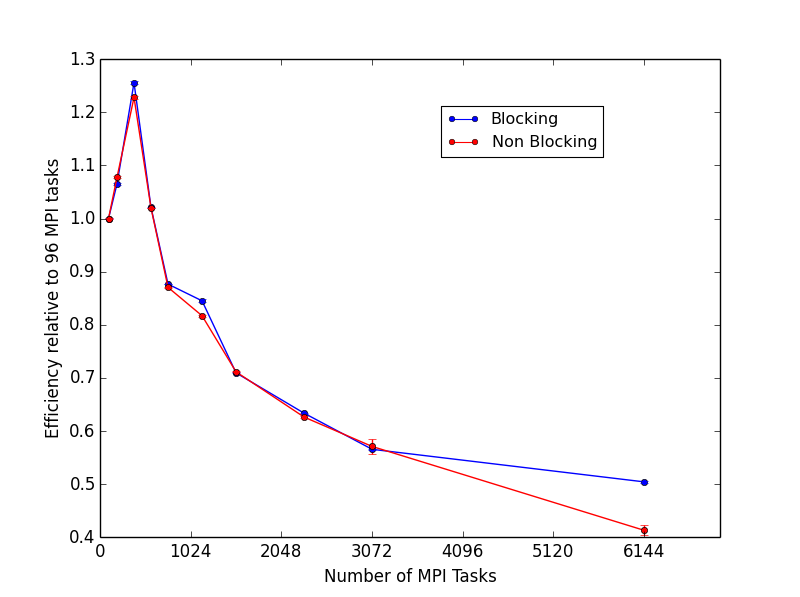}
		\caption{Parallel Efficiency calculated relative to same $T_1$ value for a system size of $384^3$. Error Bars are small, mostly not visible under points.}
		\label{fig:384_efficiency2}
	\end{figure}

	\paragraph{}
	Figure \ref{fig:384_speedups} shows the speedup of both versions of the code on up to 6,144 MPI Tasks. The speedup is very similar up to approximately $3,000$ MPI Tasks. But a large number of MPI Tasks appears to favour the blocking version. This, however, corresponds to very small subdomain sizes and a parallel efficiency below 60\%, see figure \ref{fig:384_efficiency2}, which would not be used in reality. Thus, this system does not show any significant performance differences between the non-blocking and blocking routines.
	
	\section{$768^3$ System Size} \label{sec:768_768_768}
	
	\begin{figure}[H]
		\centering
		\includegraphics[width=0.99\textwidth]{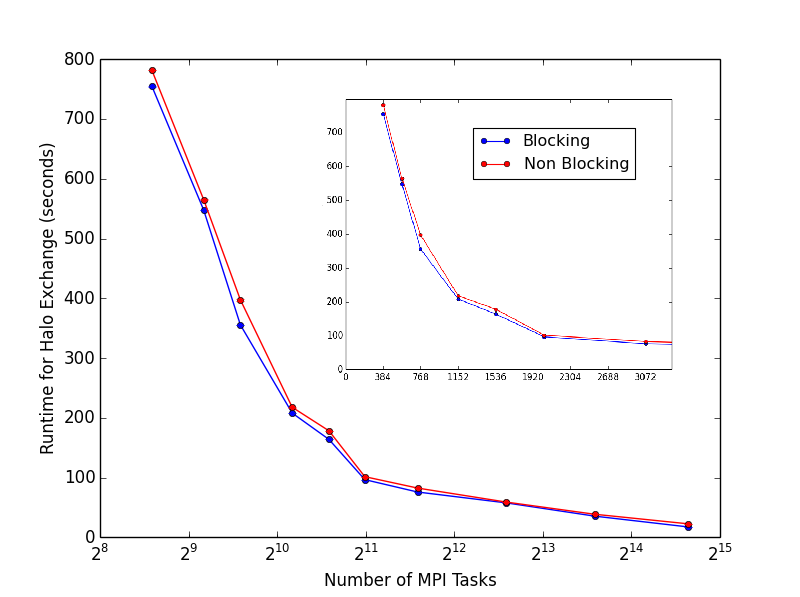}
		\caption{Time for 2,000 calls to halo exchange routine for system size of $768^3$. Error Bars are small, mostly not visible under points.}
		\label{fig:768_runtime}
	\end{figure}

\begin{figure}[H]
		\centering
		\begin{subfigure}[h]{0.49\textwidth}
			\centering
			\includegraphics[width=\textwidth]{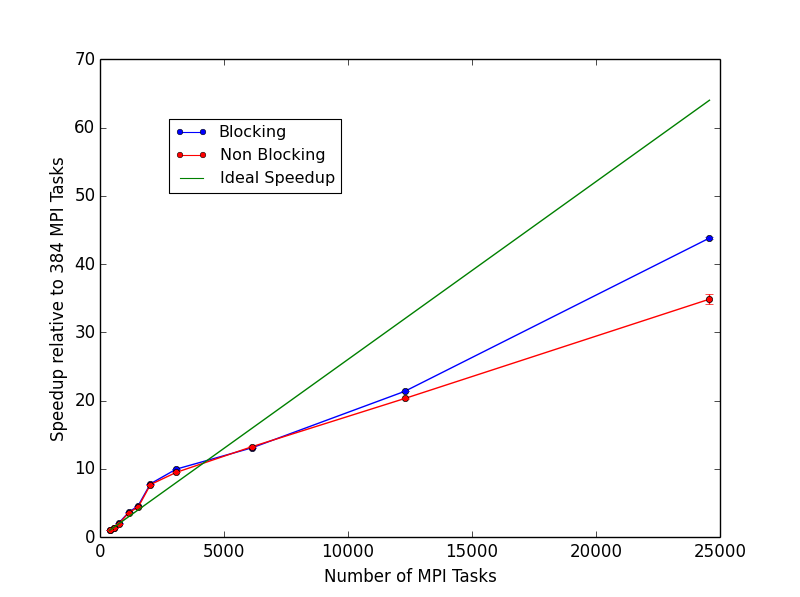}
			\caption{Speedup relative to respective $T_1$ values.}
			\label{fig:768_speedupp}
		\end{subfigure}
		\hfill
		\begin{subfigure}[h]{0.49\textwidth}
			\centering
			\includegraphics[width=\textwidth]{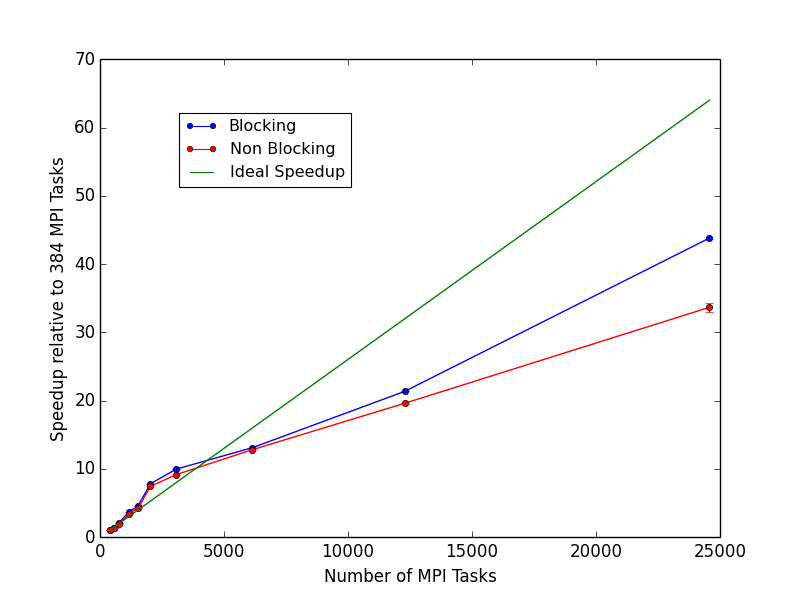}
			\caption{Speedup relative to same $T_1$ value.}
			\label{fig:768_speedup2}
		\end{subfigure} 
		\caption{Speedup for $768^3$ system size relative to 384 MPI Tasks. Error Bars are small, mostly not visible under points.}
		\label{fig:768_speedups}
	\end{figure}

	\paragraph{}
	Figure \ref{fig:768_runtime} shows the execution time of the halo exchange function for increasing number of MPI Tasks for a $768^3$ system size. The performance of both versions of the model are very evenly balanced. As before, both Speedup relative to each version's independent $T_1$ and relative to the same $T_1$ are given in figure \ref{fig:768_speedups}.
	
\begin{figure}[H]
		\centering
		\includegraphics[width=0.99\textwidth]{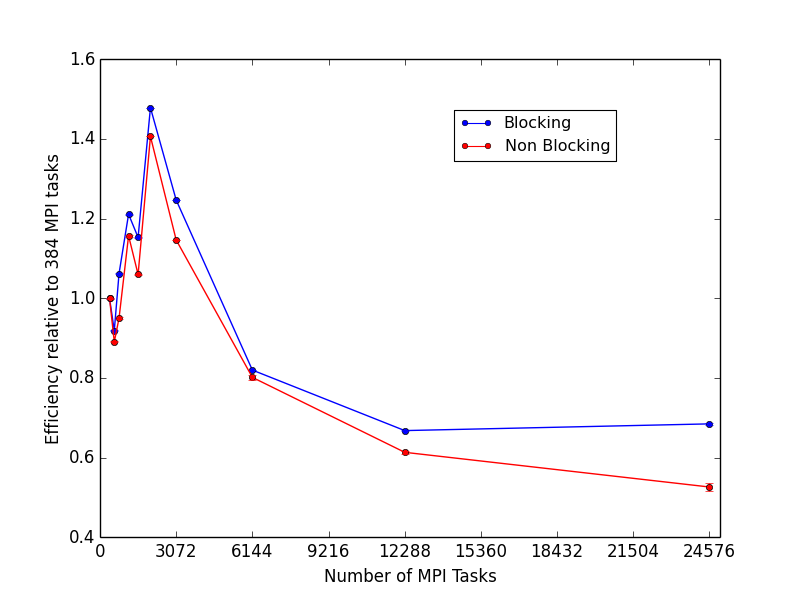}
		\caption{Efficiency calculated relative to same $T_1$ value for system size of $768^3$. Error Bars are small, mostly not visible under points.}
		\label{fig:768_efficiency2}
	\end{figure}

	\paragraph{}
	Considering the speedup in figure \ref{fig:768_speedups} the blocking version appears to scale better on a large number of MPI Tasks for large problems. However, similarly as for the $384^3$ system size, this is using an inefficient number of cores where the parallel efficiency is below 80\% as can be seen in figure \ref{fig:768_efficiency2}. Correspondingly, the runtimes for a large number of MPI tasks is relatively small, and so the speedup may seem large but in reality only be a few seconds difference in the runtime. 
	
	\begin{table}[ht]
	\caption{Subdomain size and dimensions relating to figure \ref{fig:768_runtime}.} 
	\centering 
	\begin{tabular}{c c c} 
		\hline\hline 
		Number of Cores & Subdomain Dimensions & Subdomain Size \\ [0.5ex] 
		\hline 
		384 & 128 192 192 & 4,718,592 \\
		576 & 96 128 192 &  2,359,296 \\ 
		768 & 96 96 128 &   1179648 \\
		1,152 & 64 96 128 & 786432 \\
		1,536 & 64 64 96 &  393216 \\ 
		2,034 & 48 64 96 &  294912 \\ 
		3,072 & 48 64 64 &  196608 \\
		6,144 & 32 48 48 &  73728 \\
		12,288 & 24 32 48 & 36864 \\ 
		24,576 & 24 24 32 & 18432 \\ [1ex]
		\hline 
		\label{table:768_runtime}
		\end{tabular}
	\end{table}	
	
	\paragraph{}	
	The difference in runtime on a large number of cores is nonetheless an indication of the extra cost of latency paid by the non-blocking version. This extra cost is more heavily weighting for small messages and hence why the blocking version's parallel efficiency does not diminish as quickly as the non-blocking version's.
	
	\section{Runtime Differences} \label{sec:runtime_diffs}
	
\begin{figure}[H]
		\centering
		\begin{subfigure}[h]{0.49\textwidth}
			\centering
			\includegraphics[width=\textwidth]{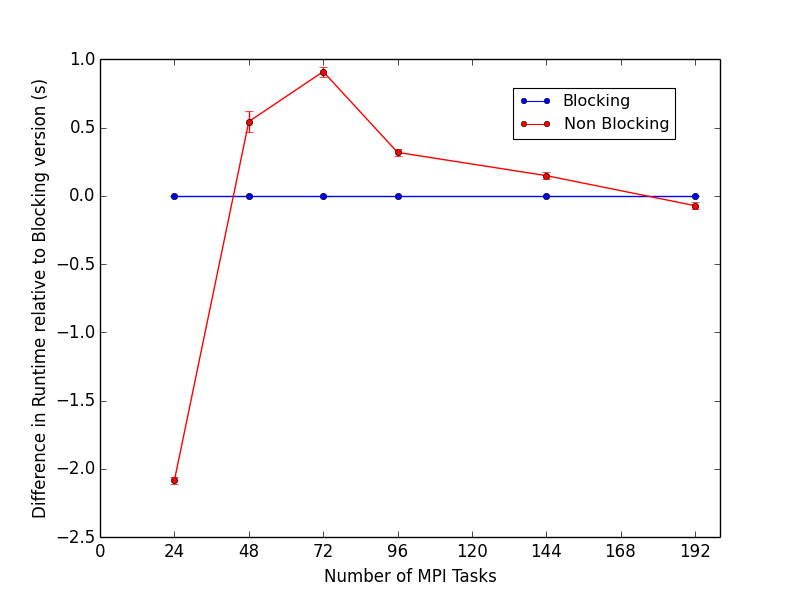}
			\caption{System Size: $96^3$.}
			\label{fig:96_diff}
		\end{subfigure}
		\hfill
		\begin{subfigure}[h]{0.49\textwidth}
			\centering
			\includegraphics[width=\textwidth]{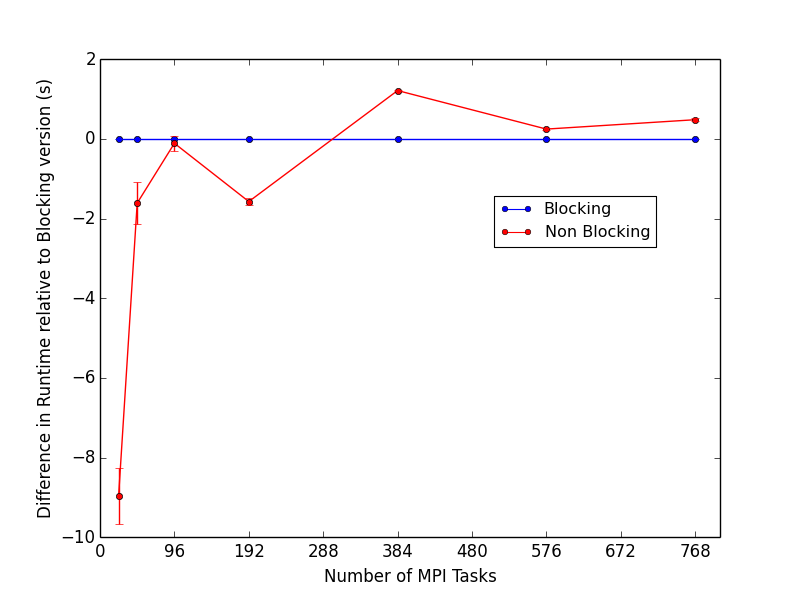}
			\caption{System Size: $192^3$.}
			\label{fig:192_diff}
		\end{subfigure} 
		\begin{subfigure}[h]{0.49\textwidth}
			\centering
			\includegraphics[width=\textwidth]{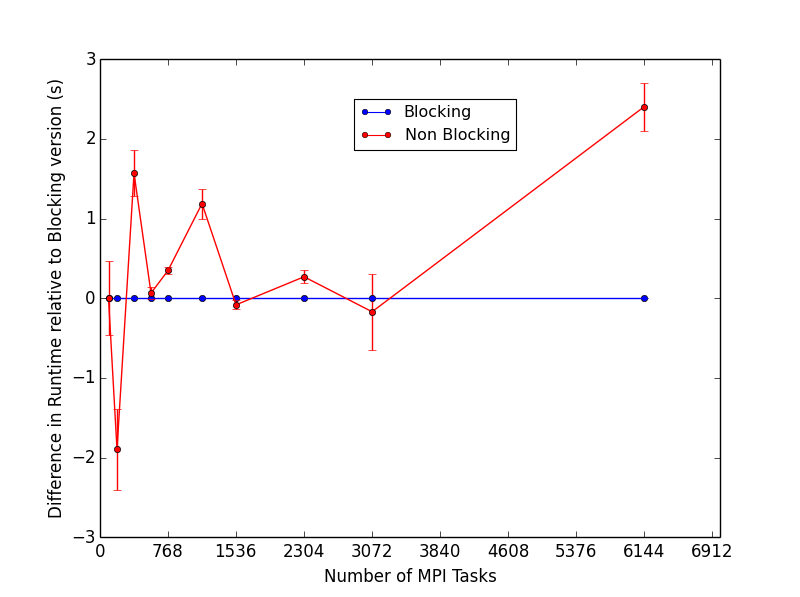}
			\caption{System Size: $384^3$.}
			\label{fig:384_diff}
		\end{subfigure}
		\hfill
		\begin{subfigure}[h]{0.49\textwidth}
			\centering
			\includegraphics[width=\textwidth]{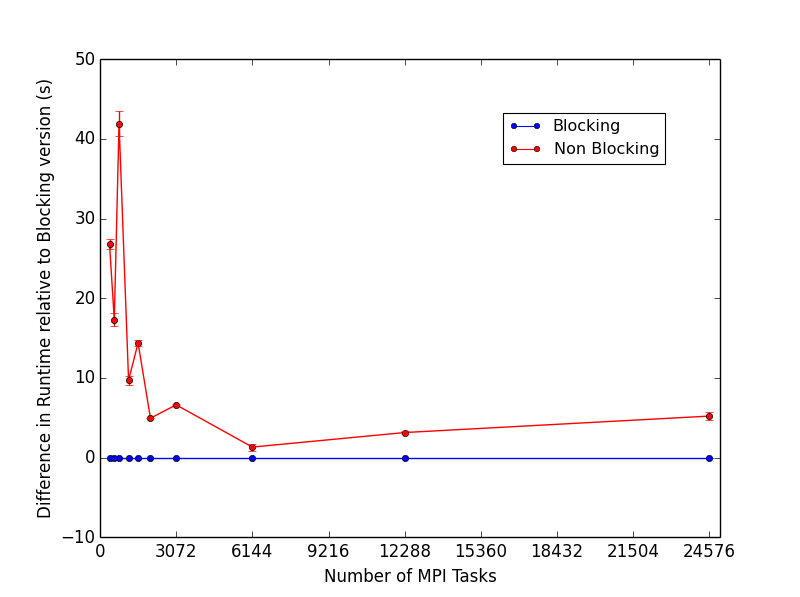}
			\caption{System Size: $768^3$.}
			\label{fig:768_diff}
		\end{subfigure}
		\caption{Runtime for 2,000 calls to the halo exchange routine of the non-blocking halo exchange routine relative to the blocking routine against Number of MPI Tasks for all strong scaling simulations.}
		\label{fig:diffs}
	\end{figure}

	\paragraph{}
	Figure \ref{fig:diffs} shows the difference in runtime for the two versions of the code relative to the blocking version's runtime for each of the above lattice sizes. When comparing this figure to the runtimes in each of the previous sections (figures \ref{fig:96_seconds}, \ref{fig:192_runtime}, \ref{fig:384_runtime} and \ref{fig:768_runtime}) it can be seen that the difference in runtime between the codes remains a small fraction of runtime, especially for a high number of MPI tasks.

	\paragraph{}
	It is particularly noticeable, in figures \ref{fig:384_diff} and \ref{fig:768_diff} that the point corresponding to the second smallest number of MPI tasks (192 for the $384^3$ system and 576 for the $768^3$, see tables \ref{table:384_runtime} and \ref{table:768_runtime}) defy the general trend of the graph. Relatively favouring the non-blocking version in each case. The aforementioned tables illustrate that these simulations correspond to the most dimensionally imbalanced subdomains in the tables. Hence, this supports the conclusion that the non-blocking version incurs less of a penalty for higher communication to work ratio than the blocking version in the case where the messages being sent are large.
	
	\paragraph{}
	Although it has been illustrated that the non-blocking version favours (relative to the blocking version) a higher communication-work ratio, this doesn't give it an advantage for a large number of cores despite the fact that the ratio of communication-work increases as the subdomain size decreases. This gives further evidence that the desynchronicity offered by the non-blocking version gives little advantage for small messages in iterations that occur over a very short time. In this case the reduced message sending latency in the blocking version is a similar weighting trade-off to desynchronisation.

\chapter{Work-Communication Overlap} \label{chapter:overlap}

\paragraph{}
As discussed in chapter \ref{chapter:development}, the call to \texttt{lb\_halo\_via\_copy()} was replaced with a call to \texttt{lb\_halo\_via\_copy\_nonblocking\_start()} and a \\ \texttt{lb\_halo\_via\_copy\_nonblocking\_end()}. The purpose of including a \texttt{start} and \texttt{end} function is that overlapping computational work, that is not dependent on the halos, may be placed between the start and end functions. A considerable amount of work in this area will, in practice, eliminate any form of blocking and be fully optimised. It is considered important to analyse the difference between the raw non-blocking halo routine and the blocking version as the results are generalised and not problem dependent which is why the rest of the report does not provided analysis of the overlapped work and communication.

\paragraph{}
This optimisation is problem dependent and will be applicable to different specific work for different kinds of lattice models. This chaper, however, gives a shallow but insightful indication of the potential optimisation offered by non-blocking communication.

\paragraph{}
This chapter considers three versions of the code: the original blocking version; the non-blocking version; and finally the non-blocking version with the \texttt{build\_update\_map} function between the \texttt{start} and \texttt{end} of the non-blocking halo routine. Timings in this benchmark are total simulation time instead of time spent in the halo exchange routine. This is because the overlap version does additional work after the halo exchange routine begins and before it ends. Thus the halo exchange routine time is a biased comparison in this case.

\begin{figure}[H]
	\centering
	\includegraphics[width=0.99\textwidth]{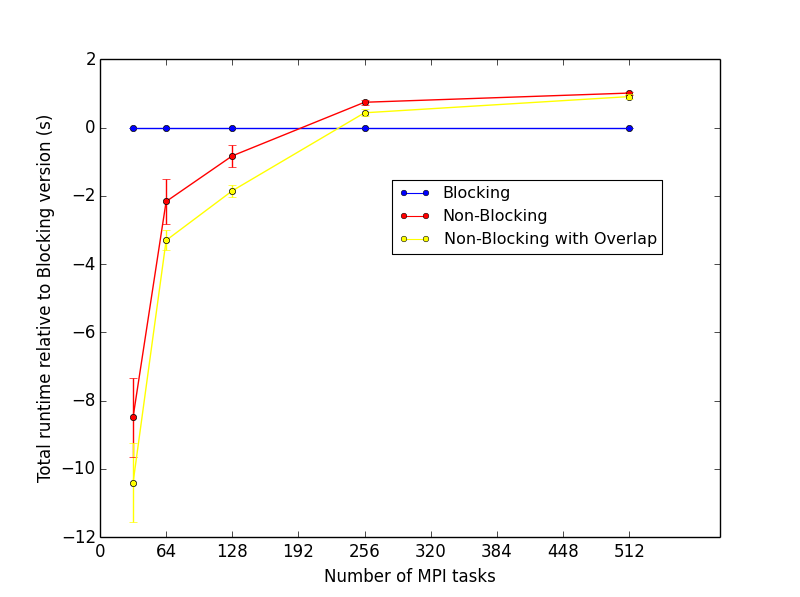}
	\caption{Total runtime for each version relative to the blocking version total runtime for system size of $256 \times 256 \times 128$. Total runtime is the time to run 1,000 time step loops. Error Bars are small, mostly not visible under points.}
	\label{fig:overlap}
\end{figure}

\paragraph{}
Figure \ref{fig:overlap} shows the runtime for each version of the code relative to the original blocking version's runtime for a system size of $256 \times 256 \times 128$.

\paragraph{}
Both non-blocking versions have an advantage over the blocking version for large subdomain sizes (corresponding to few MPI tasks). The non-blocking versions scale slightly poorer as the number of MPI tasks gets high and, correspondingly, as the subdomain sizes and the runtime gets very small. This is subject to the same discussion as in chapter \ref{chapter:strong_scaling}. It can be seen that the non-blocking version with overlap of work and communication is consistently faster than the non-blocking version which demonstrates the advantage of work-communication overlap.   

\begin{figure}[H]
	\centering
	\includegraphics[width=0.8\textwidth]{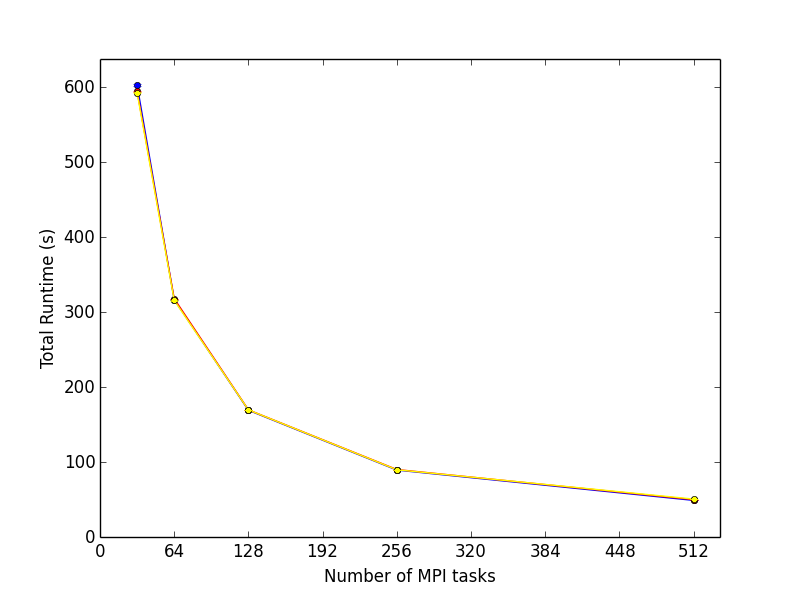}
	\caption{Runtime for system size of $256 \times 256 \times 128$. Error Bars are small, mostly not visible under points.}
	\label{fig:overlap_runtime}
\end{figure}

\begin{figure}[H]
	\centering
	\includegraphics[width=0.8\textwidth]{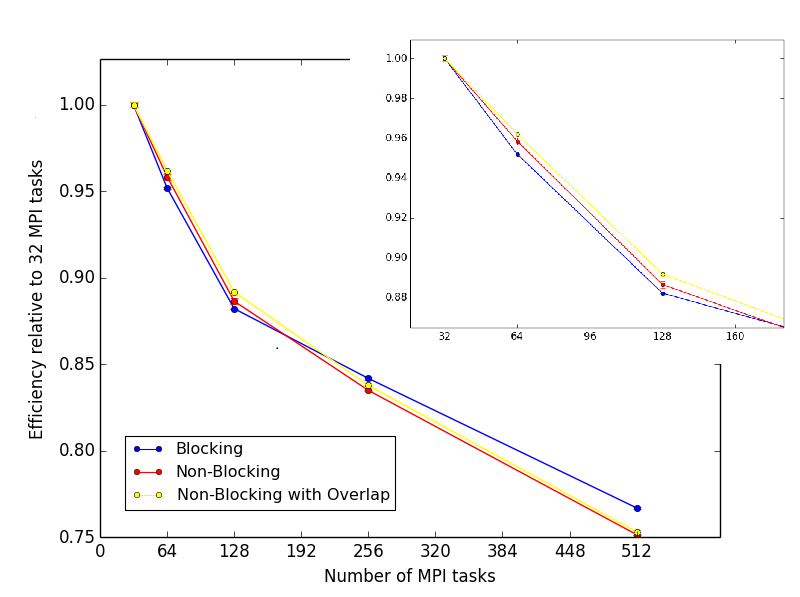}
	\caption{Parallel Efficiency for system size of $256 \times 256 \times 128$. Error Bars are small, mostly not visible under points.}
	\label{fig:overlap_efficiency}
\end{figure}

\paragraph{}
It can be seen, from figure \ref{fig:overlap_runtime} that the differences in runtime of the three different versions are a small fraction of overall runtime. Thus, for this problem the differences are insignificant. As discussed in chapter \ref{chapter:strong_scaling}, the cost of latency becomes a significant factor for a large number of MPI tasks since subdomain sizes, and therefore messages, are small. These small subdomain sizes also correspond to simulations with poor parallel efficiency as seen in figure \ref{fig:overlap_efficiency}. 

\chapter{Conclusions} \label{chapter:conclusions}

\paragraph{}
A non-blocking implementation of halo exchange has been described, designed and implemented for the lattice Boltzman code called \textit{Ludwig} \cite{ludwig}. The code was successfully implemented and tested as an alternative to the blocking version. 

\paragraph{}
A ping-pong test on ARCHER estimated the approximate Bandwidth $B$ for MPI communication. It was found to saturate at message sizes greater than $\approx 0.5 MBytes$. The Effective Bandwidth at this size is $B_{eff} \approx 300 MBytes/s$. The maximum Effective Bandwidth is also correspondingly found to have a limiting factor of $\approx 360 MBytes/s$. This test was conducted between two MPI Tasks from cores on separate computing nodes.  

\paragraph{Cubic Subdomains}
The performance of the non-blocking version relative to the blocking version always increases for larger subdomains. This performance advantage has been seen to be very significant between subdomain sizes $58^3$ and $88^3$. The rate of updates per core is maximised for both version of the code for subdomain sizes between $58^3$ and $64^3$. The subdomain size that optimises the rate of updates per core can be used to determine a sensible subdomain size for weak scaling. The non-blocking version favours weak scaling with large subdomain sizes which is useful for large simulations if hardware resources are limited. Both versions have a very similar performance for smaller, and more usually used, subdomain sizes with neither one demonstrating a significant advantage over the other. 

\paragraph{Non-Cubic Subdomains}
The purpose of examining Non-Cubic subdomains was to determine the direct influence of a higher communication-work ratio on the performance of each version of the code, while excluding the factor of message size. The results showed that the non-blocking version incurs less of a performance penalty in simulations where subdomains are less dimensionally balanced than its blocking counterpart. Thus implying that the performance of the non-blocking version is better relative to the blocking version when imposed with a higher communication to work ratio. 

\paragraph{}
The non-blocking version has (approximately) an, at least, equal performance to the blocking version for all simulations. There is no significant difference in performance for a large number of MPI tasks, especially in the range of standard simulations which will have subdomain sizes between $24^3$ and $48^3$.  

\paragraph{}
The advantages that the non-blocking version has over the blocking version are generally higher than the disadvantage due to the latency of a greater number of messages. For small message sizes the advantage of desychronicity (which has shown to result in better performance for high communication-work ratios) offered by the non-blocking version is an equally weighting trade off to reduced latency in the blocking version. This results in both versions of the code having very similar performances when scaled onto a large number of cores. Thus the non-blocking version is a good alternative with some advantages and no substantial disadvantages over its blocking counterpart and so may be preferably used even without any work-communication overlap.

\paragraph{}
Work-communication overlap in the non-blocking version of the code always improves performance relative to the raw non-blocking version. The magnitude of the advantage is problem dependent.

	\section{Further Study} \label{sec:further_study}
	
	\paragraph{}
	Further code development, apart from changing the work-communication overlap depending on the problem to be solved, may involve unpacking buffers containing the halo data once the relevant receive has completed. This may offer further improvement over waiting for all receives before unpacking buffers however was considered to require too much more development time to be included in this project. A full implementation of this method should give a similar performance to the non-blocking version presented in this report when work has been overlapped with communication.

	\newpage
	\section{Acknowledgements}
		I would like extend my gratitude to my project supervisor: Dr Kevin Stratford for many useful meetings and emails throughout the project.

\begin{appendices}

	\chapter{3D Images}
	
	\begin{figure}[H]
		\centering
		\includegraphics[width=0.9\textwidth]{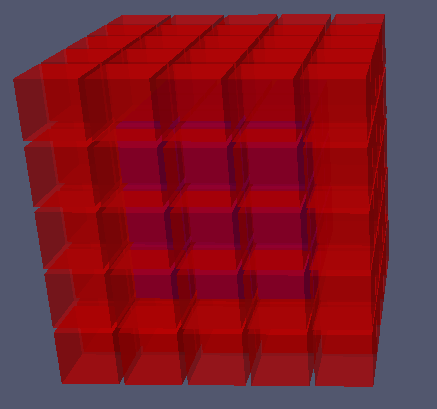}
		\caption{Subdomain with complete set of halo data. Subdomain size is $3 \times 3 \times 3 = 27$ lattice sites..}
		\label{fig:complete_halo}
	\end{figure}	
	
	\chapter{Graphs}
	
	\begin{figure}[H]
		\centering
		\begin{subfigure}[h]{0.49\textwidth}
			\centering
			\includegraphics[width=\textwidth]{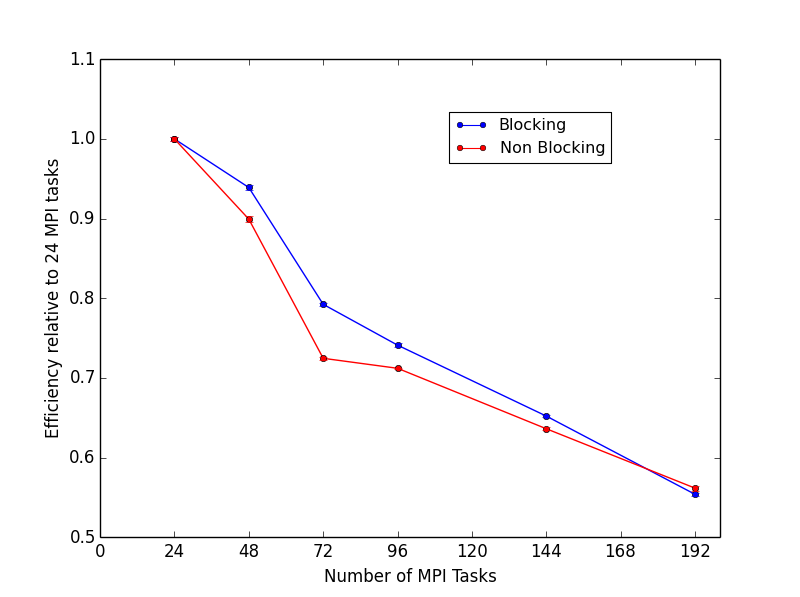}
			\caption{System Size: $96^3$.}
			\label{fig:96_efficiency2_appen}
		\end{subfigure}
		\hfill
		\begin{subfigure}[h]{0.49\textwidth}
			\centering
			\includegraphics[width=\textwidth]{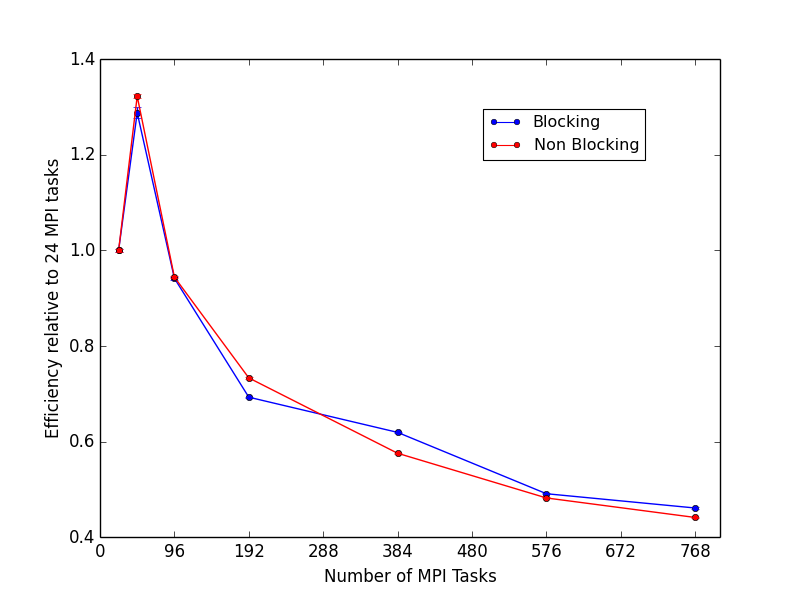}
			\caption{System Size: $192^3$.}
			\label{fig:192_efficiency2_appen}
		\end{subfigure} 
		\begin{subfigure}[h]{0.49\textwidth}
			\centering
			\includegraphics[width=\textwidth]{384_efficiency2.png}
			\caption{System Size: $384^3$.}
			\label{fig:384_efficiency2_appen}
		\end{subfigure}
		\hfill
		\begin{subfigure}[h]{0.49\textwidth}
			\centering
			\includegraphics[width=\textwidth]{768_efficiency2.png}
			\caption{System Size: $768^3$.}
			\label{fig:768_efficiency2_appen}
		\end{subfigure}
		\caption{Parallel efficiencies of the Blocking and Non-Blocking versions for the system sizes shown. These are relative to the same rubtime on the minimum number of cores ($T_1$).}
		\label{fig:efficiencies2_appen}
	\end{figure}

\end{appendices}

\end{document}